\documentclass[preprintnumbers,nofootinbib,
superscriptaddress,amsmath,amssymb,floatfix,prd,10pt]{revtex4}

\usepackage[english]{babel}
\usepackage{graphicx}
\usepackage{bm}
\usepackage{epsf}
\usepackage{amssymb,amsmath}

\usepackage{color}

\usepackage[normalem]{ulem}  

\newcommand{\be}{\begin{equation}}
\newcommand{\ee}{\end{equation}}
\newcommand{\bea}{\begin{eqnarray}}
\newcommand{\eea}{\end{eqnarray}}
\newcommand{\non}{\nonumber\\}

\renewcommand{\vec}[1]{\bm{#1}}

\begin{document}

\title{Dissipation triggers dynamical two-stream instability}

\author{Nils Andersson}
\email{n.a.andersson@soton.ac.uk}
\affiliation{Mathematical Sciences and STAG Research Centre, University of Southampton, Southampton SO17 1BJ, United Kingdom}
\author{Andreas Schmitt}
\email{a.schmitt@soton.ac.uk}
\affiliation{Mathematical Sciences and STAG Research Centre, University of Southampton, Southampton SO17 1BJ, United Kingdom}

\date{1 November 2019}

\begin{abstract}
Two coupled, interpenetrating fluids suffer instabilities beyond certain 
critical counterflows. For ideal fluids, an energetic instability occurs at the point where a sound mode inverts its direction due to the counterflow, while dynamical instabilities only occur at larger relative velocities. Here we discuss two relativistic fluids, one of which is dissipative. Using linearized hydrodynamics, we show that in this case the energetic instability turns dynamical,  i.e., there is an exponentially growing mode, and this exponential growth only occurs in the presence of dissipation. This result is general and does not rely on an underlying microscopic theory. It can be applied to various two-fluid 
systems for instance in the interior of neutron stars. We also point out that under certain circumstances the two-fluid system exhibits a mode analogous to the $r$-mode 
in neutron stars that can become unstable for arbitrarily small values of the counterflow. 
\end{abstract}

\maketitle


\section{Introduction and conclusions}
\label{sec:intro}

Two-fluid (or multi-fluid) systems are realized in various settings. They can 
be created in the laboratory, for instance through Bose-Fermi mixtures of cold atomic gases \cite{2014Sci...345.1035F,2015PhRvL.115z5303D,PhysRevLett.117.145301}. They may also appear in neutron star
cores, where neutrons, protons, and electrons, possibly together with hyperons, can be described as a multi-fluid or even a multi-superfluid system \cite{1992ApJ...395..250G,Comer:2002dm,Gusakov:2009kc,Glampedakis:2010ec}. Besides the core of a neutron star, the inner crust hosts a complicated system of superfluid neutrons (and electrons) immersed in a lattice of ions \cite{Chamel:2008ca,Schmitt:2017efp}. This is somewhat comparable to an atomic superfluid in an optical lattice, which shows instabilities analogous to the two-fluid systems discussed here \cite{PhysRevA.64.061603}. 
Also, any single superfluid at finite temperature is a two-fluid system, in which the condensate and the collective excitations act as independent fluids \cite{tisza38,landau41,Alford:2012vn}. 
Two-fluid models have also been proposed in the 
context of the quark-gluon plasma to account for strongly and weakly coupled sectors
\cite{Kurkela:2018dku} and for dark matter being gravitationally coupled to normal matter in neutron stars \cite{Leung:2011zz,Xiang:2013xwa} and quark stars \cite{Mukhopadhyay:2015xhs}. 

Here we are interested in two miscible relativistic fluids which are coupled to each other but have independent velocity fields. We shall work in a hydrodynamic formalism, and our main result does not 
depend on the microscopic theory from which 
the equation of state and transport coefficients can be obtained. (We will use a specific equation of state to illustrate our results.) The validity of our two-fluid hydrodynamics relies on the following assumptions: the mean free paths from collisions of the microscopic constituents of each fluid among themselves are much smaller than the system size (that's why each of the 
components can be treated as a fluid), and the mean free path from inter-collisions, i.e., between constituents of one fluid with the other, is larger than those from intra-collisions (that's why the system does not simply become a single fluid), but still smaller than the system size (otherwise the fluids would not interact). 

The main motivation for this study is as follows: If the two coexisting fluids move relative to each other, an instability can be expected at a certain critical velocity, termed two-stream instability or counterflow instability \cite{Buneman:1959zz,1963PhRvL..10..279F,2001AmJPh..69.1262A}. Here we are only discussing homogeneous systems, i.e., the fluids  interpenetrate each other and we do not consider interfaces 
between the fluids, which also exhibit counterflow instabilities 
\cite{Livescu_2011}. Instabilities of relativistic ideal  fluids in this homogeneous scenario were discussed systematically in Refs.\ \cite{2004MNRAS.354..101A,Haber:2015exa}. It was found that {\it dynamical} instabilities only occur at very large velocities, where there is already an {\it energetic} instability, which 
sets in at a smaller counterflow. Here, dynamical instability refers to an exponentially increasing mode with a certain growth time. Since we shall only work with linearized hydrodynamic equations, this growth time is valid for the initial stage of the instability before nonlinear effects become important.  An energetic instability, on the other hand, refers to a negative energy, but 
there is not necessarily a growth time associated with it. The negative energy indicates that the system wants to relax to a state of lower energy. For this instability to turn into an actual unstable process, i.e., for it to turn dynamical, the coupling to an environment is needed, for instance by dissipative processes. Demonstrating and explaining this effect for the two-stream instability in a two-fluid system is the main goal of this paper. In our 
hydrodynamic approach, the energetic instability is associated with a change of direction of a sound mode from upstream to downstream due to an increasing counterflow between the fluids. In a more microscopic approach, this flip over
is accompanied by the energy of a quasiparticle excitation turning negative \cite{Haber:2015exa}. This is completely analogous to Landau's original argument for the instability of a superflow, which thus is an example for an energetic instability. The relation between energetic and dynamical instabilites is also discussed for 
instance in Refs.\ \cite{PhysRevA.64.061603,2007PhRvA..76f3607R,YU20181231}. 
Another example of an energetic instability is a negative canonical 
energy of rotating fluids \cite{1978ApJ...221..937F}, and we shall point out further intriguing parallels of our observations to rotational instabilities in neutron stars below.

We consider a two-fluid system in which one of the fluids is dissipative. Dissipation is taken into account by adding terms of first order in space-time derivatives to the stress-energy tensor. In a single fluid, this gives rise to three dissipation channels, characterized by  three transport coefficients:  heat 
conductivity, shear viscosity, and bulk viscosity. Two-fluid systems allow for more coefficients. This is known for instance from a superfluid at finite temperature, which has one dissipative and one non-dissipative component, resulting in three bulk viscosity coefficients \cite{khala,Mannarelli:2009ia}.
 For our purpose it is sufficient to consider only the dissipative coefficients that are already present in the absence of the second fluid and thus ignore additional dissipative effects due to the counterflow between the fluids. We also do not discuss two-fluid systems where both fluids would be dissipative in the 
absence of the other fluid. Moreover, for our main result, we shall  restrict ourselves to zero temperature, i.e., only viscosity, not 
heat conduction, will play a role. In this sense, we are constructing 
the "minimal" scenario which shows our main result: 
dissipation triggers 
a dynamical instability exactly at the critical velocity at which, in the absence of dissipation, an energetic instability would occur.  

It is known that in first-order hydrodynamics unphysical 
instabilities occur even for a single fluid \cite{Hiscock:1985zz}, which are cured by going to second order \cite{Hiscock:1987zz} (or possibly, as recently suggested \cite{Kovtun:2019hdm,Bemfica:2019knx}, by generalizing the common first-order formulations by Eckart and Landau/Lifshitz). We shall therefore 
start with a careful discussion of a single relativistic dissipative fluid (in an arbitrary rest frame) and identify the unphysical modes before moving on to two coupled fluids. Then, by working in the
rest frame of the dissipative fluid and at zero temperature, we ensure that the modes with unphysical instabilities do not mix with the modes we are interested in. As a further confirmation that the two-stream instabilities considered here are not artifacts of first-order 
relativistic hydrodynamics, we shall also discuss the non-relativistic limit, where no unphysical instabilities occur and where the same conclusion is reached. The observation of the coincidence of an energetic instability and a dynamical instability due to dissipative effects has also been made in 
the study of a holographic superfluid \cite{Amado:2013aea}. 

Our system may be considered as a toy version of a finite-temperature superfluid. In that case, 
dissipative and non-dissipative components are connected by a microscopic equation of state, and the dissipative 
component is identical to the entropy current. In the present study we keep the coupling between the fluids completely general, allowing 
also for entrainment -- a nontrivial mixing of the currents and their conjugate momenta.  In the future it would be interesting to repeat the calculation using an actual equation of state for a finite temperature superfluid, for instance along the lines of Refs.\ \cite{Alford:2013koa,Schmitt:2013nva}. The generality of our setup  allows us to decide whether the non-dissipative fluid behaves as a  normal fluid or a superfluid. In a superfluid the hydrodynamic equations are supplemented by a Josephson equation which shuts off certain modes that exist in a normal fluid. Our main result does not depend on whether or not this constraint is taken into account, i.e., it holds for both normal fluids and superfluids. 

The results of our study are mostly of general value, but they also have concrete implications. For 
instance, it has been suggested that two-stream instabilities may provide the trigger 
for a collective unpinning of vortices from the ion lattice in the inner crust of a neutron star, which is necessary to explain sudden jumps in the rotation frequency, so-called pulsar glitches, see for instance Refs.\  \cite{Peralta:2006um,Chamel:2012pk,Andersson:2012iu,Haskell:2015jra}. By comparing the estimated lag between the rotating superfluid and the crust from glitch data with the regions in parameter space where the two-stream instability sets in 
it has been argued that the instability only affects very large multipole moments \cite{2004MNRAS.354..101A}. This estimate, however, was based on the ideal fluid approximation. Our results show that 
dynamical instabilities in realistic systems set in at lower 
counterflow velocities, which suggests that the instability may operate at lower multipole moments and thus on larger length scales. 
Therefore, further studies which connect our general results to concrete systems such as neutron star crusts, but also to two-fluid systems in the laboratory,  would be of great interest. 

\begin{figure} [t]
\begin{center}
\includegraphics[width=0.8\textwidth]{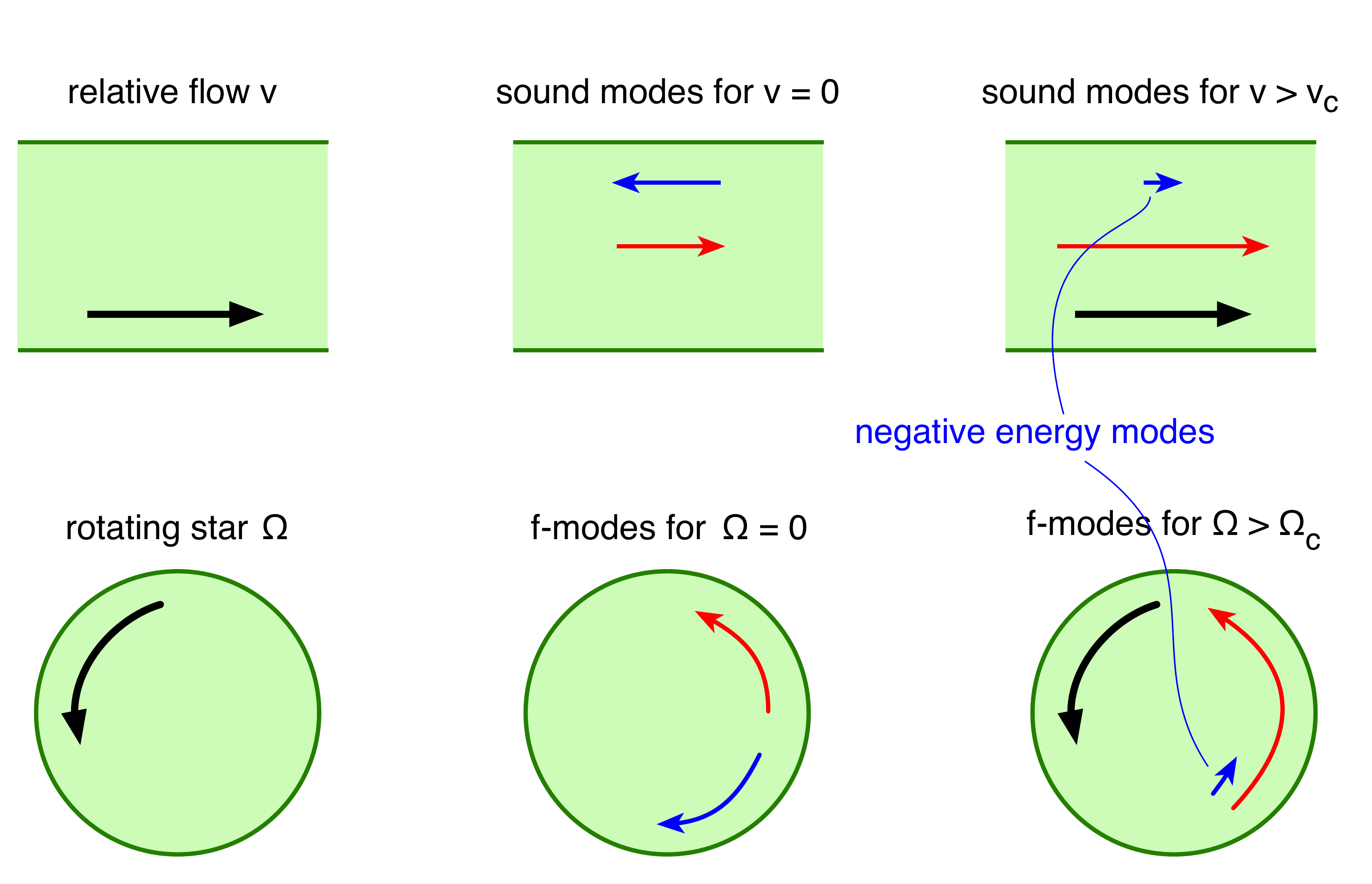}
\caption{Analogy between the two-stream instability (upper row, discussed in this paper),
and the CFS instability in neutron stars, here schematically shown for $f$-modes (lower row). {\it Left column:} The thick (black) arrow indicates the relative velocity $v$ of the second fluid, viewed in the rest frame of the first fluid, or the
rotation of the star with frequency $\Omega$, viewed from an inertial frame. {\it Middle column:} In the absence of a relative motion of the fluids or a rotation of the star, 
a given mode can propagate with the same speed in two opposite directions, indicated by two thin (red and blue) arrows of the same length. {\it Right column:} At a sufficiently large counterflow between the fluids or a sufficiently large rotation frequency of the star, one of the modes flips over. This point marks the onset of an energetic instability. (In the astrophysics literature, instabilities indicated by negative energies are often referred to as "secular instabilities" \cite{1977ApJ...213..497H,Friedman:1978hf}.)   
If the system is now allowed to interact with its environment by dissipative effects (both cases) or losing angular
momentum through the emission of gravitational waves (star), the energetic instability turns dynamical. 
}
\label{fig:secular}
\end{center}
\end{figure}   

Putting the result in a wider context, we note that our observation is not confined to a linear relative flow between two fluids. In Fig.\ \ref{fig:secular} we show the analogy between our system and the case of a rotating star where certain oscillation modes suffer the so-called Chandrasekhar-Friedman-Schutz (CFS) instability \cite{PhysRevLett.24.611,1978ApJ...221..937F}, which manifests itself 
in the emission of gravitational waves. In this figure we have sketched the mechanism 
for $f$-modes -- "fundamental" modes of the star 
whose instabilities are driven by dissipation or gravitational radiation
\cite{1991ApJ...373..213I,Gaertig:2011bm}. However, the phenomenologically more interesting instability is predicted for $r$-modes \cite{Andersson:1997xt} because it sets in at much smaller rotation frequencies. (For a recent review about $f$-mode and $r$-mode instabilities and their significance for gravitational wave emission of isolated neutron stars see Ref.\ \cite{Glampedakis:2017nqy}.) The $r$-modes only exist in a rotating star and, in the absence of any damping mechanism, become unstable at infinitesimally small rotation frequencies. Translated to our two-fluid system, $r$-modes would find an analogy if there is a mode that only propagates in the presence of a counterflow and which, if dissipation is switched on, becomes immediately unstable. We demonstrate the 
existence of such a mode if entrainment between the two fluids is taken into account. 
Intriguingly, this observation provides us with a system that is much simpler than the rotating star coupled to gravitational waves and yet shows qualitatively 
the same instability. This suggests that the $r$-mode instability 
is a more general phenomenon than previously thought and not only a consequence of the complicated details of rotating stars and general relativity. 
It would be interesting to exploit the analogy to the two-fluid system further in the future.


\section{Single fluid}
\label{sec:single}

\subsection{Setup}

To prepare the two-fluid calculation and establish our notation, we start by discussing a single dissipative fluid in first-order hydrodynamics.
We are interested in oscillatory modes, which will be obtained in the usual way by evaluating the hydrodynamic equations to linear order in the fluctuations. With the two-fluid application in mind, we shall discuss these modes in a general 
rest frame. This generalizes the well-known results of Ref.\ \cite{Hiscock:1985zz}, which are obtained in the rest frame of the fluid. The present formalism for relativistic dissipative fluids has of course been frequently applied in the literature, see for instance Refs.\ \cite{Mannarelli:2009ia,Kapusta:2011gt,Kovtun:2012rj,Strickland:2014pga}.

Our starting point is the conserved current $j^\mu$ and the stress-energy tensor 
$T^{\mu\nu}$, 
\begin{subequations}
\bea
j^\mu &=& nv^\mu \, , \\[2ex]
T^{\mu\nu} &=& T^{\mu\nu}_{\rm ideal} + T^{\mu\nu}_{\rm diss} \,,   
\eea
\end{subequations}
where $n$ is the number density in the rest frame of the fluid, and $v^\mu$ is the four-velocity, which obeys $v_\mu v^\mu=1$. We work in Minkowski space with metric convention $g^{\mu\nu} = (1,-1,-1,-1)$. 
We have adopted a formulation in which the current does not receive dissipative corrections. This is referred to as the Eckart frame \cite{PhysRev.58.919} (in contrast to the Landau/Lifshitz frame 
formulation \cite{Landau:Fluid}). Up to first order in derivatives, the ideal and dissipative 
contributions are 
\begin{subequations}
\bea
T^{\mu\nu}_{\rm ideal} &=& \epsilon v^\mu v^\nu - P\Delta^{\mu\nu} \, , \\[2ex]
T^{\mu\nu}_{\rm diss} &=& -\tau \Delta^{\mu\nu} + q^\mu v^\nu+q^\nu v^\mu+\tau^{\mu\nu} \, , \label{Tdiss}
\eea
\end{subequations}
where $\epsilon$ and $P$ are energy density and pressure in the rest frame of the fluid, and 
$\Delta^{\mu\nu} \equiv g^{\mu\nu} - v^\mu v^\nu$.
Here, $q^\mu$ is four-transverse to $v^\mu$ (i.e., $q\cdot v \equiv q_\mu v^\mu = 0$), and $\tau^{\mu\nu}$ is four-transverse to $v^\mu$, traceless, and symmetric. The explicit form of the dissipative contribution in terms of the 
transport coefficients is given by  
\begin{subequations}\label{tauq}
\bea
\tau &=& -\zeta\partial\cdot v \, , \\[2ex]
q^\mu &=& \kappa \Delta^{\mu\gamma}[\partial_\gamma T-T(v\cdot\partial)v_\gamma] \, , \\[2ex]
\tau^{\mu\nu} &=& \eta\Delta^{\mu\gamma}\Delta^{\nu\delta}\left(\partial_\delta v_\gamma+\partial_\gamma v_\delta - \frac{2}{3}g_{\gamma\delta} \partial\cdot v\right) \, ,
\eea
\end{subequations}
with temperature $T$, shear viscosity $\eta$, bulk viscosity $\zeta$, and heat conductivity $\kappa$. For the explicit calculation it is useful to write the heat and shear terms in the stress-energy tensor as 
\begin{subequations} \label{qvqv}
\bea
q^\mu v^\nu+q^\nu v^\mu &=& 2\kappa\left[v^{(\mu}\partial^{\nu)}T-v^\mu v^\nu(v\cdot\partial)T-Tv^{(\mu}(v\cdot\partial)v^{\nu)}\right] \, , \\[2ex]
\tau^{\mu\nu}
&=& 2\eta \left[\partial^{(\mu} v^{\nu)} -v^{(\mu}(v\cdot \partial)v^{\nu)} -\frac{1}{3}\Delta^{\mu\nu}\partial\cdot v\right] \, , 
\eea
\end{subequations} 
where we have used $v^\gamma\partial^\mu v_\gamma=0$ and abbreviated $A^{(\mu}B^{\nu)} \equiv \frac{1}{2}(A^\mu B^\nu + A^\nu B^\mu)$. 

The hydrodynamic equations are given by the conservation of current, energy, and momentum,
\be \label{conserve}
\partial_\mu j^\mu = \partial_\mu T^{\mu\nu} = 0 \, .
\ee
It is instructive to compute the entropy production, which also
serves as a consistency check for the signs of the dissipative contributions. Treating the chemical potential $p$ and the temperature $T$ as our independent thermodynamic variables\footnote{In Sec.\ \ref{sec:twofluids} we shall denote the chemical potentials in the laboratory frame (in which both fluids have a nonzero velocity) by $\mu_1$ and $\mu_2$, while the chemical potentials in the rest frames of the 
fluids will be denoted by $p_1$ and $p_2$. Hence, for consistency, we 
employ the slightly unusual notation $p$ for the chemical potential already here,
although in this section the chemical potential in the laboratory frame $\mu$ never appears.} 
and denoting the entropy density by $s$, we compute  
\begin{subequations}\label{djTa}
\bea
\partial_\mu j^\mu &=& n\partial\cdot v+\frac{\partial n}{\partial p}(v\cdot\partial)p+\frac{\partial s}{\partial p}(v\cdot\partial) T \, , \\[2ex]
\partial_\mu T^{\mu\nu} &=& p v^\nu \partial_\mu j^\mu +p n(v\cdot\partial)v^\nu +sT[v^\nu\partial\cdot v+(v\cdot\partial)v^\nu] \non[2ex]
&&+\left[\left(n+T\frac{\partial n}{\partial T}\right) v^\nu v\cdot\partial -n\partial^\nu\right] p + \left[\left(s+T\frac{\partial s}{\partial T}\right) v^\nu v\cdot\partial -s\partial^\nu\right] T
+\partial_\mu T_{\rm diss}^{\mu\nu} \,, \label{dT3}
\eea
\end{subequations}
where 
all derivatives with respect to $p$ are taken at fixed $T$ and vice versa. We have used the thermodynamic relations $dP=ndp+sdT$, $d\epsilon=pdn+Tds$, $n= \frac{\partial P}{\partial p}$, $s= \frac{\partial P}{\partial T}$, and $\epsilon+P=pn+sT$, which imply $dn=\frac{\partial n}{\partial p}dp+\frac{\partial s}{\partial p}dT$ and $ds=\frac{\partial n}{\partial T}dp+\frac{\partial s}{\partial T}dT$. 
With $\partial_\mu j^\mu=0$ and the definition of the 
dissipative part of the stress energy tensor (\ref{Tdiss}), the contraction of Eq.\ (\ref{dT3}) with the four-velocity yields 
\be \label{Tsmu}
v_\nu\partial_\mu T^{\mu\nu} = T\partial_\mu(sv^\mu)+\tau\partial\cdot v+\partial\cdot q + v_\nu(v\cdot\partial)q^\nu + v_\nu\partial_\mu\tau^{\mu\nu} \, , 
\ee
where the transversality 
of $q^\mu$ has been employed. 
The terms on the right-hand side are rewritten with the help 
of the explicit forms of the dissipative terms (\ref{tauq}), 
\bea
\tau\partial\cdot v &=& -\frac{\tau^2}{\zeta} \, , \qquad 
v_\mu(v\cdot\partial)q^\mu = \frac{q_\mu q^\mu}{\kappa T}-\frac{(q\cdot\partial) T}{T} \, , \qquad v_\nu\partial_\mu\tau^{\mu\nu} = -\frac{\tau_{\mu\nu}\tau^{\mu\nu}}{2\eta} \, .
\eea
With these expressions and $\partial_\mu T^{\mu\nu}=0$ Eq.\ (\ref{Tsmu}) becomes 
\be \label{ds}
T\partial_\mu s^\mu = \frac{\tau^2}{\zeta}-\frac{q_\mu q^\mu}{\kappa T} +\frac{\tau_{\mu\nu}\tau^{\mu\nu}}{2\eta} \, ,
\ee
where we have defined the entropy current 
\be
s^\mu \equiv  sv^\mu +\frac{q^\mu}{T} \, . 
\ee
Each of the terms in the entropy production (\ref{ds}) is 
non-negative provided that  $\eta,\zeta,\kappa>0$, as it should be. For instance, in the rest frame of the fluid we can easily write the 
entropy production terms in the form of squares, 
\bea
\tau^2 = \zeta^2(\nabla\cdot\vec{v})^2 \, , \qquad q_\mu q^\mu = -\kappa^2(\nabla T+T\partial_0\vec{v})^2 \, , \qquad \tau_{\mu\nu}\tau^{\mu\nu} &=& \eta^2\left(\partial_i v_j+\partial_j v_i-\frac{2}{3}\delta_{ij}\nabla\cdot\vec{v}\right)^2 \, , 
\eea
where we have kept derivatives of the three-velocity $\vec{v}$, but set $\vec{v}=0$ otherwise. The three-velocity is defined via $v^\mu =\gamma(1,\vec{v})$, where $\gamma=1/\sqrt{1-v^2}$ is the usual Lorentz factor with $v=|\vec{v}|$. Here and in the following, the components of the three-velocity are denoted by $v_i$, which should not be confused with the 
spatial components of the four-velocity (which in general have an additional factor $\gamma$).
One can also write the entropy production (\ref{ds}) in terms of the so-called thermodynamic
forces $\partial_\mu v_\nu$ and $[\partial_\mu T-T(v\cdot\partial)v_\mu]/T$ and the corresponding thermodynamic fluxes $\pi^{\mu\nu} = \tau\Delta^{\mu\nu}-\tau^{\mu\nu}$ and
$q^\mu$,  
\be \label{entprod}
\partial_\mu s^\mu = -\frac{\pi^{\mu\nu}\partial_\mu v_\nu+q^\mu[\partial_\mu T-T(v\cdot\partial)v_\mu]/T}{T} \,, 
\ee
which can be compared to the non-relativistic form, see for instance Ref.\ \cite{Schmitt:2017efp} and Eq.\ (\ref{Entropy}) in 
our discussion of the non-relativistic limit. 

In order to compute the sound modes of the system we introduce harmonic
fluctuations 
in chemical potential, temperature, and three-velocity about their equilibrium values, 
\begin{subequations} \label{fluc}
\bea
p(\vec{x},t) &=& p + \delta p \, e^{i(\omega t- \vec{k}\cdot\vec{x})} \, , \\[2ex]
T(\vec{x},t) &=& T + \delta T \, e^{i(\omega t- \vec{k}\cdot\vec{x})} \, , \\[2ex]
\vec{v}(\vec{x},t) &=& \vec{v} + \delta \vec{v} \, e^{i(\omega t- \vec{k}\cdot\vec{x})} \, . 
\eea
\end{subequations}
We insert this ansatz into Eqs.\ (\ref{djTa}) to obtain the non-dissipative contributions [we omit the exponential $e^{i(\omega t- \vec{k}\cdot\vec{x})}$, which multiplies every term on the right-hand sides]
\begin{subequations}
\bea
\partial_\mu j^\mu &=& i\gamma n\left[\frac{\omega_v}{n}\left(\frac{\partial n}{\partial p}\delta p + \frac{\partial s}{\partial p}\delta T\right) -(\vec{k}-\gamma^2\omega_v\vec{v})\cdot\delta\vec{v}\right] \, , \label{ideal1}\\[2ex]
\partial_\mu T^{\mu 0}_{\rm ideal} &=&  
i\gamma^4(w+sT)\omega_v\vec{v}\cdot\delta\vec{v}-i\gamma^2sT\vec{k}\cdot\delta\vec{v}  \non[2ex]
&&+i\left[n(\gamma^2\omega_v-\omega)+\gamma^2T\frac{\partial n}{\partial T}\omega_v\right]\delta p 
+i\left[s(\gamma^2\omega_v-\omega)+\gamma^2T\frac{\partial s}{\partial T}\omega_v\right]\delta T \, , \label{ideal2} \\[2ex]
\partial_\mu T^{\mu i}_{\rm ideal} &=& i\gamma^4(w+sT)\omega_vv_i\vec{v}\cdot\delta\vec{v}-i\gamma^2sTv_i\vec{k}\cdot\delta\vec{v}+iw\gamma^2\omega_v\delta v_i  \non[2ex]
&&+i\left[n(\gamma^2\omega_v v_i-k_i)+\gamma^2 T\frac{\partial n}{\partial T} \omega_vv_i\right]\delta p 
+i\left[s(\gamma^2\omega_v v_i-k_i)+\gamma^2 T\frac{\partial s}{\partial T} \omega_vv_i\right]\delta T \, .\label{ideal3}
\eea
\end{subequations}
Here we have used $\partial_\mu j^\mu=0$, have defined the enthalpy density 
\be
w=p n +sT \, , 
\ee
and have abbreviated 
\be
\omega_v \equiv \omega - \vec{k}\cdot\vec{v} \, .
\ee
To compute the dissipative contributions we linearize in the fluctuations $\delta p$, $\delta T$, $\delta\vec{v}$. In this approximation, and using Eqs.\ (\ref{Tdiss}) and (\ref{qvqv}), we can write 
\bea
\partial_\mu T_{\rm diss}^{\mu\nu} &\simeq& \left(\zeta+\frac{\eta}{3}\right)
[\partial^\nu-v^\nu(v\cdot\partial)]\partial\cdot v +\eta[\Box-(v\cdot\partial)^2]v^\nu \non[2ex]
&&+\kappa\Big\{v^\nu[\Box-2(v\cdot\partial)^2]+(v\cdot\partial)\partial^\nu\Big\} T-\kappa T\left[(v\cdot\partial)^2 v^\nu+v^\nu(v\cdot\partial)(\partial\cdot v)\right]
\, ,
\eea
where $\Box\equiv \partial_\mu\partial^\mu$. 
This yields the spatial and temporal components (again omitting the exponential)
\begin{subequations} \allowdisplaybreaks\label{DissExpl}
\bea
\partial_\mu T^{\mu 0}_{\rm diss} &\simeq& 
 -\gamma^3\left[\eta(\omega^2-\gamma^2\omega_v^2-k^2)+\left(\zeta+\frac{\eta}{3}\right)\omega_v(\omega-\gamma^2\omega_v) -2\kappa T\gamma^2\omega_v^2\right] \vec{v}\cdot\delta\vec{v} \non[2ex]
&& +\gamma\left[\left(\zeta+\frac{\eta}{3}\right)(\omega-\gamma^2\omega_v)-\kappa T\gamma^2\omega_v\right]\vec{k}\cdot\delta\vec{v} -\kappa\gamma\Big[\omega^2-\gamma^2\omega_v^2-k^2
+\omega_v(\omega-\gamma^2\omega_v)\Big]\delta T \, , \\[2ex]
\partial_\mu T^{\mu i}_{\rm diss} &\simeq& 
-\gamma^3\left[\eta v_i (\omega^2-\gamma^2\omega_v^2-k^2)+\left(\zeta+\frac{\eta}{3}\right)\omega_v (k_i-\gamma^2\omega_v v_i)-2\kappa T v_i \gamma^2\omega_v^2\right] \vec{v}\cdot\delta\vec{v} \non[2ex]
&& +\gamma\left[\left(\zeta+\frac{\eta}{3}\right)(k_i-\gamma^2\omega_v v_i)-\kappa T\gamma^2\omega_v v_i\right]
 \vec{k}\cdot\delta\vec{v} -\kappa\gamma\Big[v_i(\omega^2-\gamma^2\omega_v^2-k^2) +\omega_v(k_i-\gamma^2\omega_v v_i)\Big]\delta T \non[2ex]
 &&-\gamma\Big[\eta(\omega^2-\gamma^2\omega_v^2-k^2)-\kappa T\gamma^2\omega_v^2\Big]\delta v_i \, .
\eea
\end{subequations}

\subsection{Sound modes at $T=0$ in the rest frame of the fluid}

As a warm-up, let us start with the simplest case: zero temperature, 
$T=s=\kappa=0$, and vanishing fluid velocity, $\vec{v} = 0$. In this case, the temporal 
component of the energy-momentum conservation, $\partial_\mu T^{\mu 0}=0$,  is automatically fulfilled. We contract the spatial components $\partial_\mu T^{\mu i}$ by $k_i$ and $\delta_{ij}-\hat{k}_i\hat{k}_j$  to obtain 
(three-)longitudinal and (three-)transverse components with respect to the direction of propagation of the mode. Together with the current conservation, this yields
\begin{subequations} \label{djdT0}
\bea
0&=& \omega\delta p-p c^2 \vec{k}\cdot\delta\vec{v} \, , \label{dj10} \\[2ex]
0&=& - k^2\delta p + \left(p\omega-ik^2\frac{4\eta+3\zeta}{3n} \right)\vec{k}\cdot\delta\vec{v}  \, , \label{dTl0}\\[2ex]
0&=& \left(p\omega-ik^2\frac{\eta}{n}\right)\delta \vec{v}_\perp \, , \label{dTt0} 
\eea
\end{subequations} 
where $\delta v_{\perp,i}=(\delta_{ij}-\hat{k}_i\hat{k}_j)\delta v_j$, and
\be \label{csquared}
c^2=\frac{n}{p}\left(\frac{\partial n}{\partial p}\right)^{-1}
\ee
is the squared sound speed. 
From Eqs.\ (\ref{dj10}) and (\ref{dTl0}) we find the longitudinal modes (where $\delta \vec{v}_\perp=0$) 
\be \label{pmc}
\omega = \pm k\sqrt{c^2-k^2\Gamma_0^2}+ik^2\Gamma_0 \simeq \pm ck + ik^2\Gamma_0 +{\cal O}(k^3) \, ,  
\ee
where we have denoted the attenuation constant by 
\be \label{Gamma0}
\Gamma_0 \equiv \frac{4\eta+3\zeta}{6w} \, . 
\ee
(Here, at zero temperature, $w=p n$.) The attenuation constant has units of time. Note, however, that the 
damping time scale from the imaginary part in the 
$k^2$ contribution is set by $(k^2 \Gamma_0)^{-1}$ and thus depends on the wavenumber $k$. 
The modes (\ref{pmc}) are damped since ${\rm Im} \, \omega = k^2\Gamma_0>0$. 
Imaginary parts of $\omega$ with the opposite sign would indicate unstable modes. 

From Eq.\ (\ref{dTt0}) we find one transverse mode, where $\delta \vec{v}_\perp\neq 0$ but $\delta p=\vec{k}\cdot\delta\vec{v}=0$,  with 
\be \label{Tp0}
\omega = \frac{i\eta}{w} k^2 \, .
\ee
This is a purely diffusive mode which does not propagate.

\subsection{Sound modes at nonzero $T$ in general frame}

Let us now discuss the general single-fluid case. In particular we allow for a nonzero velocity $\vec{v}$. For a single fluid without boundaries such as walls of a capillary and without coupling to a second fluid, this nonzero velocity should of course not lead to any new physics since it is nothing but a Lorentz boost of the whole system. Nevertheless, the calculation will turn out to be useful as a preparation for the two-fluid system. 

The equations $\partial_\mu j^\mu=\partial_\mu T^{\mu\nu}=0$ are 5 scalar equations for the 5 scalar variables $\delta p$, $\delta T$, $\delta\vec{v}$. Since we allow for a general frame, we now have a second three-vector in the problem, namely the fluid velocity  $\vec{v}$, in addition to the wavevector $\vec{k}$. We align the $z$-axis with 
$\vec{k}$ and the $y$-axis such that the fluid velocity $\vec{v}$ is in the 
$y$-$z$ plane. Instead of the 
components $\delta v_x$, $\delta v_y$, $\delta v_z$ we work with $\delta v_x$, $\vec{k}\cdot\delta\vec{v}$, 
and $\vec{v}\cdot\delta\vec{v}$, and instead of $\partial_\mu T^{\mu i}=0$ we work with  $\partial_\mu T^{\mu x}=v_i \partial_\mu T^{\mu i}=k_i \partial_\mu T^{\mu i}=0$. 
The equation $\partial_\mu T^{\mu x}=0$ decouples because it is the only equation where the transverse fluctuation $\delta v_x$ appears (and no other 
fluctuations). The prefactor of $\delta v_x$ is a quadratic polynomial in $\omega$ with the solutions 
\be \label{omegai}
\omega = -i\frac{w+2i\gamma(\eta+\kappa T)\vec{k}\cdot\vec{v} \pm \sqrt{w^2+4iw\eta \vec{k}\cdot\vec{v}/\gamma +4\eta\kappa T k^2(1-v^2\cos^2\theta)
+4\eta^2k^2v^2(1-\cos^2\theta)}}{2\gamma(\kappa T+\eta v^2)} \, ,
\ee
where $\theta$ is the angle between $\vec{k}$ and $\vec{v}$. Since the
velocity oscillations of these modes are transverse to both $\vec{k}$ and $\vec{v}$ we term them $T^\pm$. Their expansion up to second order in $k$ is shown in Table \ref{table0}. We see that by setting $v=T=0$ in the mode $T^+$ we recover
Eq.\ (\ref{Tp0}). In other words, the purely diffusive mode $T^+$ now becomes 
propagating in the boosted fluid. This seems to be a trivial observation, but we shall see that these modes become
particularly interesting in the case of two coupled fluids with entrainment, where they can couple to other modes, see Sec.\ \ref{sec:rmode}. The mode $T^-$ has an unphysical instability, which is 
an artifact of first-order hydrodynamics \cite{Hiscock:1985zz}. 
Going to second-order
hydrodynamics cures this unphysical instability \cite{Hiscock:1987zz}\footnote{In the notation of Ref.\ \cite{Hiscock:1987zz}, the modes shown in Table \ref{table0} are, from top to bottom, $L_4^+$, $L_4^-$, $T_4$, $T_3$, $L_5$, $L_2$, while in the presence of the second-order terms additional modes, termed $T_1$, $T_2$, $L_1$, $L_3$, are found.}.
The mode $T^-$ did not appear in the previous subsection, where we 
worked in the limit $v=T=0$ from the beginning. It is useful to keep this in mind for our discussion of the two-fluid system, where we shall work in the zero-temperature limit and in the rest frame of the dissipative fluid, such that the unphysical instability plays no role.

\begin{table}[t]
\begin{center}
\begin{tabular}{|c||c|c|c|} 
\hline
\rule[-1.5ex]{0em}{4ex} 
$\;\;$Mode$\;\;$ &  1 & $k$ & $k^2$ \\[1ex] \hline\hline
\rule[-1.5ex]{0em}{6ex} $L^+$& 0&$+c$& $\displaystyle{i\Gamma_0 + i\kappa[\mbox{see Eq.\ (\ref{Gammaezk})}]}$  \\[2ex] \hline
\rule[-1.5ex]{0em}{6ex} 
$L^-$&0&$-c$& $\;\;\displaystyle{i\Gamma_0 + i\kappa[\mbox{see Eq.\ (\ref{Gammaezk})}]}\;\;$ \\[2ex] \hline
\rule[-1.5ex]{0em}{6ex} 
$T^+$ & 0& $v\cos\theta$ & $\displaystyle{\frac{i\eta(1-v^2\cos^2\theta)}{\gamma w}} $  \\[2ex] \hline
\rule[-1.5ex]{0em}{6ex} 
$T^-$ & $\;\;\displaystyle{-\frac{iw}{\gamma(\eta v^2+\kappa T)}}\;\;$ & $\;\;\displaystyle{v\cos\theta\frac{\eta(2-v^2)+\kappa T}{\eta v^2+\kappa T}}\;\;$  & $\displaystyle{-\frac{i\eta(1-v^2\cos^2\theta)}{\gamma w}}$ \\[2ex] \hline
\rule[-1.5ex]{0em}{6ex} 
& 0 & $v\cos\theta$ & $\;\;i\kappa[\mbox{see Eq.\ (\ref{omegv})}]\;\;$  \\[2ex] \hline
\rule[-1.5ex]{0em}{6ex} 
& $\displaystyle{-\frac{iw}{\kappa T}}$& 0 & $\displaystyle{-2i\Gamma_0-i\kappa[\mbox{see Eq.\ (\ref{wk0})}]}$ \\[2ex] \hline
\end{tabular}
\caption{Coefficients of the single-fluid modes in a low-momentum expansion 
of $\omega(\vec{k})$ up to $k^2$, with $v$-dependence shown only where the expression is short enough, i.e., for $L^\pm$ and the mode in the 6th row only the $v=0$ result is given. 
The modes in the 4th and 6th row show an unphysical instability in first-order hydrodynamics.  
The modes of the bottom two rows only exist at nonzero temperature, and  
play no role in our discussion of the two-fluid system, where we restrict ourselves 
to zero temperature. }
\label{table0}
\end{center}
\end{table}

The remaining 4 coupled equations give rise to a matrix of coefficients whose determinant is a polynomial of degree 6 in $\omega$, which 
factorizes into a quadratic and a quartic polynomial. The quadratic 
polynomial gives exactly the same dispersion as for the modes 
$T^\pm$ (\ref{omegai}). The four solutions 
of the quartic polynomial are obviously complicated in general.
Two of them only exist at nonzero temperatures, i.e., they are not connected to any of the modes found in the previous subsection. They are \be \label{wk0}
\omega = -\frac{iw}{\kappa T}-i\left[2\Gamma_0+\frac{\kappa n^2}{T w^2}\frac{p^2\frac{\partial n}{\partial p}+p T\left(\frac{\partial n}{\partial T}
+\frac{\partial s}{\partial p}\right)+T^2\frac{\partial s}{\partial T}}{\frac{\partial n}{\partial p}\frac{\partial s}{\partial T}-  
\frac{\partial n}{\partial T}\frac{\partial s}{\partial p}}\right]k^2 + \ldots 
\ee
(for brevity we only give the dispersion of this mode for 
$v=0$) and
\bea \label{omegv}
\omega = v\cos\theta\, k +\frac{i \kappa n^2\sqrt{1-v^2}(1-v^2\cos^2\theta)}{T\left[n^2\frac{\partial s}{\partial T}+s^2\frac{\partial n}{\partial p}
-ns\left(\frac{\partial n}{\partial T}+\frac{\partial s}{\partial p}\right)\right]}\, k^2+\ldots  \, .
\eea
The mode (\ref{wk0}) shows an (unphysical) instability, just like the mode $T^-$. The two modes (\ref{wk0}) and (\ref{omegv}) 
are included in Table \ref{table0} in the last two rows. 

Finally, there are two stable, damped modes that are generalizations 
to nonzero temperature of the two modes 
 (\ref{pmc}). They have the form $\omega = \pm c k+i\Gamma k^2 +{\cal O}(k^3)$ with 
(again we only give the $v=0$ result)
\begin{subequations}
\bea
c^2 &=& \frac{n^2\frac{\partial s}{\partial T}+s^2\frac{\partial n}{\partial p}-ns\left(\frac{\partial n}{\partial T}+\frac{\partial s}{\partial p}\right)}{w\left(\frac{\partial n}{\partial p}\frac{\partial s}{\partial T}-  
\frac{\partial n}{\partial T}\frac{\partial s}{\partial p}\right)}  = \frac{\partial P}{\partial \epsilon} \, , \label{csq} \\[2ex]
\Gamma &=& \Gamma_0+ \frac{\kappa n^2}{2w^3c^2T}\frac{\left[s\left(p\frac{\partial n}{\partial p}+T\frac{\partial n}{\partial T}\right)-n\left(p\frac{\partial s}{\partial p}+T\frac{\partial s}{\partial T}\right)\right]
\left[T\left(s\frac{\partial s}{\partial p}-n\frac{\partial s}{\partial T}\right)+p\left(s\frac{\partial n}{\partial p}-n\frac{\partial n}{\partial T}\right)\right]}
{\left(\frac{\partial n}{\partial p}\frac{\partial s}{\partial T}- \frac{\partial n}{\partial T}\frac{\partial s}{\partial p}\right)^2}  \, . \hspace{0.5cm}\label{Gammaezk}
\eea
\end{subequations}
Here, all derivatives with respect to the chemical potential $p$ are taken at fixed temperature $T$ and vice versa,
while the derivative with respect to $\epsilon$ in Eq.\ (\ref{csq}) is taken at fixed entropy per particle; see for instance appendix E of Ref.\ \cite{BitaghsirFadafan:2018uzs} for a derivation of this change of thermodynamic variables.  
At zero temperature, the sound speed (\ref{csq}) reduces to the  expression (\ref{csquared}). We term these two modes $L^\pm$, see first 
two rows of Table \ref{table0}. 

\begin{figure} [t]
\begin{center}
\hbox{\includegraphics[width=0.5\textwidth]{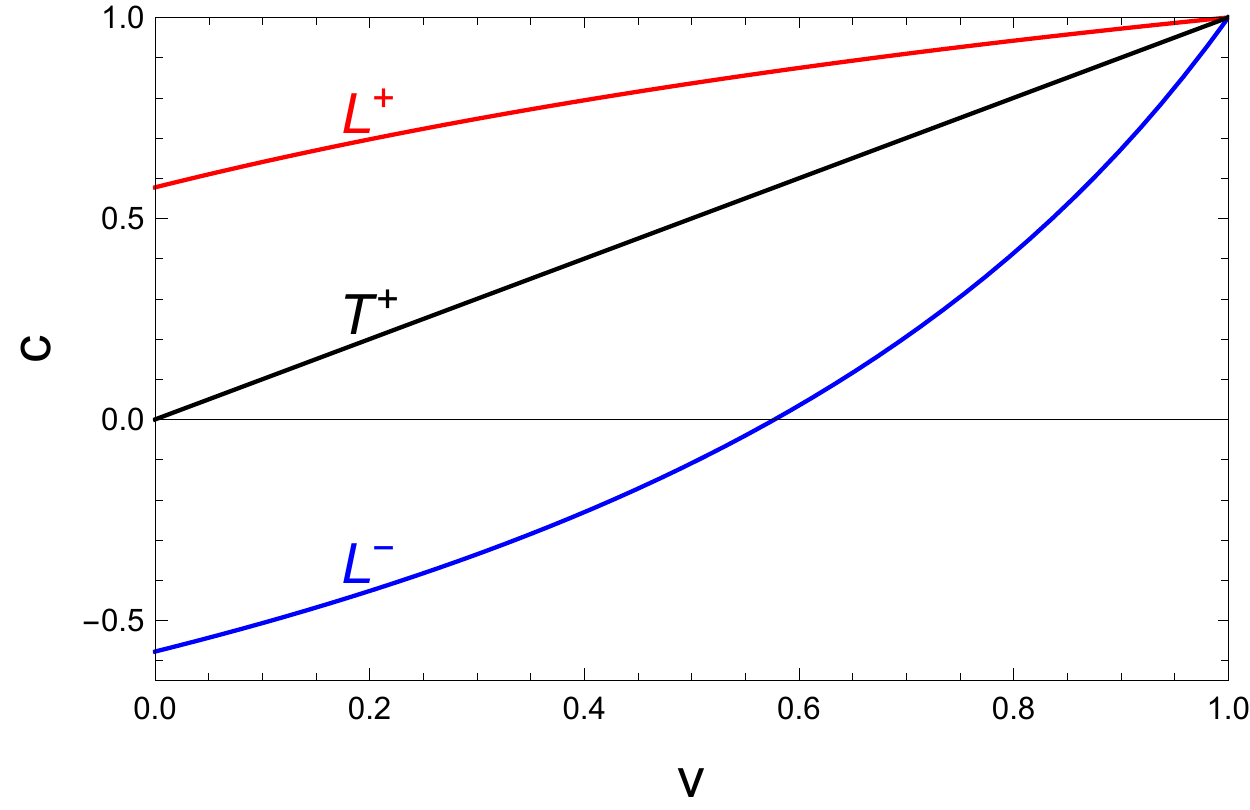}\includegraphics[width=0.5\textwidth]{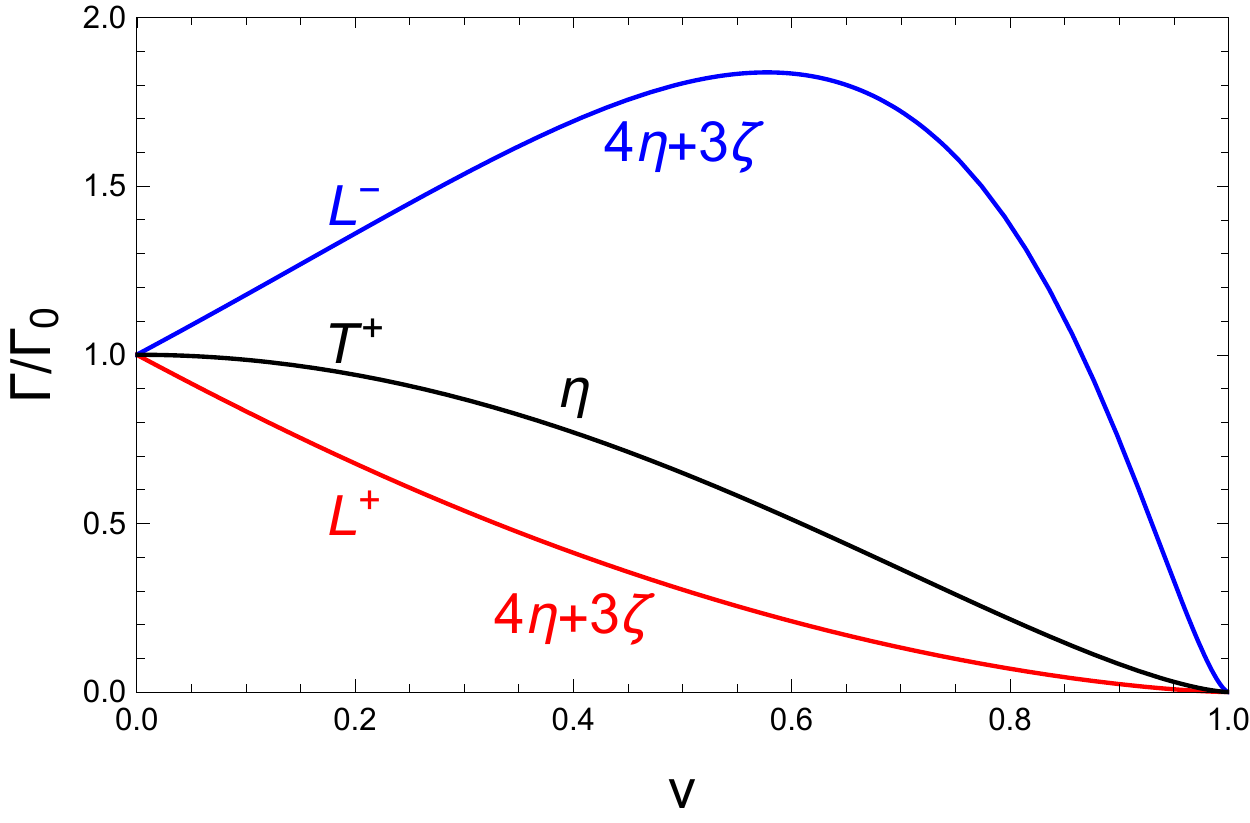}}
\caption{Single-fluid, zero-temperature sound speeds $c$ (left) and attenuation $\Gamma$ 
(right) of the three modes $L^\pm$, $T^+$ (see first three rows of Table \ref{table0}) as a function of the boost 
velocity $v$. Velocities are given in units of the speed of light, and the attenuation constant is given in units of $\Gamma_0$ from Eq.\ (\ref{Gamma0})
for $L^\pm$ and in units of $\eta/w$ for $T^+$.  
We have set $c(v=0)=1/\sqrt{3}$, but any other value of $c<1$ yields the same qualitative picture.}
\label{fig:singleUpDown}
\end{center}
\end{figure} 

In Fig.\ \ref{fig:singleUpDown} we plot the 3 stable modes that exist at $T=0$, i.e., $L^+$, $L^-$, and $T^+$, as a function of $v$. It is useful to understand this figure before moving to the two-fluid system. The modes $L^+$ and $L^-$ correspond to 
sound waves propagating in opposite directions, parallel and anti-parallel to the fluid velocity $\vec{v}$. (Equivalently, we can reinstate the dependence on $\theta$ and plot a single mode, say $L^+$, 
for $\theta=0$ and $\theta=\pi$.) For an external observer, the mode $L^+$ is 
sped up by the fluid flow, while the mode $L^-$ is slowed down until it flips over at $v=c$. This inversion of direction is not accompanied by any dynamical instability in the case of a single fluid:  the attenuation is positive for all fluid velocities. This is expected since the fluid is stable at rest and isolated and thus no dynamical instabilities should be created by simply moving the fluid. The figure also 
illustrates, as already mentioned above, that the mode $T^+$  
only appears to propagate if the fluid moves, and we see that 
its damping time is determined solely by the shear viscosity $\eta$. 
This is in contrast to the modes $L^\pm$, whose damping times
are given by a certain linear combination of shear and bulk viscosity. 

These results can also be obtained by applying a Lorentz boost
$\Lambda^{\mu\nu}$ with 3-velocity $\vec{v}$ to $k^\mu = (\omega,\vec{k})$ with $\omega = ck+i\Gamma k^2$. Then, denoting the transformed 4-momentum by 
$k'^\mu = (\omega',\vec{k}')$ with $\omega' = c'k'+i\Gamma' k'^2$, 
the relation $\Lambda^{\mu\nu} k_\nu= k'^\mu$ yields the transformed 
sound speed $c'$ and the transformed attenuation $\Gamma'$. For boosts in the direction of the wave propagation we find 
\bea \label{cGamLor}
c' = \frac{c+v}{1+vc} 
\, , \qquad 
\Gamma' = \Gamma \, \frac{(1-v^2)^{3/2}}{(1+vc)^3} \, .
\eea
Obviously, the transformed speed $c'$ is simply given by the relativistic addition of velocities. One can check that the velocity dependence of the 
attenuation shown in the right panel of Fig.\ \ref{fig:singleUpDown} is
indeed given by $\Gamma'$.

\section{Two fluids}
\label{sec:twofluids}

\subsection{Setup}

We set up the two-fluid system by first introducing the currents and stress-energy tensor for two ideal, coupled fluids at zero temperature. Here we follow the 
formalism used for instance in Refs.\ \cite{Carter:1995if,Comer:2002dm,Andersson:2006nr,Alford:2012vn,Haber:2015exa}. As a second step, we shall then add the dissipative terms from the previous section for one of the fluids.

As our independent hydrodynamic variables we choose the 
conjugate momenta $p_n^\mu$, where the index $n=1,2$ labels the 
two fluids. We denote the components of the conjugate momenta 
by $p^\mu_n = (\mu_n,\vec{p}_n)$, where the temporal components $\mu_n$ are the chemical potentials measured in the laboratory frame, and the spatial components are related to the three-velocities $\vec{v}_n$ by
$\vec{p}_n=\mu_n\vec{v}_n$ (neither here nor anywhere in the following do we use $n$ as a summation index). The chemical potentials in the
rest frames of the fluids are $p_n=(p_{n,\mu}p_n^\mu)^{1/2}=\mu_n(1-v_n^2)^{1/2}$. Conjugate momenta and 
four-velocities are thus related by\footnote{The notation is 
slightly ambiguous since non-bold letters are used for the modulus 
of the three-velocity $v_n=|\vec{v}_n|$ and for the modulus of the 
four-momentum $p_n=(p_{n,\mu}p_n^\mu)^{1/2}$. The conjugate momenta are the only four-vectors for which we employ this notation, and 
the modulus of $\vec{p}_n$ never appears; therefore, this notation should not cause any confusion.}
\be \label{vp}
v_n^\mu = \frac{p_n^\mu}{p_n} \, .
\ee
In a superfluid, temporal and spatial 
components of the conjugate momentum are not independent since 
the conjugate momentum can be written as the gradient of a scalar field, $p^\mu = \partial^\mu \psi$. This ensures the irrotationality of 
$\vec{p}$ (in a nonrelativistic superfluid $\vec{v}$ is irrotational, but this is not true in general). From a microscopic point of view, $\psi$ is the phase of a condensate, for instance a condensed scalar field in a bosonic theory \cite{Alford:2012vn} or the Cooper-pair
condensate in a fermionic superfluid. 

The central quantity that encodes the microscopic physics is the so-called generalized pressure $\Psi$. At zero temperature, it depends on the Lorentz scalars that can be constructed from the conjugate momenta, $\Psi=\Psi(p_1^2,p_2^2,p_{12}^2)$, where we have abbreviated $p_{12}^2\equiv p_1\cdot p_2$.
The conserved currents are obtained by 
\be
j^\mu_n = \frac{\partial \Psi}{\partial p_{n,\mu}} \, .
\ee
Consequently, we can write 
\begin{subequations} \label{j1j2}
\bea
j_1^\mu &=& {\cal B}_1 p_1^\mu + {\cal A}\, p_2^\mu \, , \\[2ex]
j_2^\mu &=& {\cal A}\, p_1^\mu + {\cal B}_2 p_2^\mu \, , 
\eea
\end{subequations}
where 
\be \label{BA}
{\cal B}_n \equiv 2\frac{\partial \Psi}{\partial p_n^2} \, , \qquad 
{\cal A} \equiv \frac{\partial \Psi}{\partial p_{12}^2} \, .
\ee 
In general, the currents are not proportional to their own conjugate momentum, 
but, if ${\cal A}\neq 0$, receive a contribution from the conjugate momentum of the 
other fluid. This effect
is called entrainment or Andreev-Bashkin effect \cite{1976JETP...42..164A}, and ${\cal A}$ is the entrainment coefficient. In the present formalism entrainment arises if the generalized pressure depends on $p_{12}^2$. Of course, even in the absence of entrainment the two fluids can be coupled to each other, 
for instance through terms in the pressure proportional to $p_1p_2$. 
We shall set up the hydrodynamic equations in full generality, including entrainment terms. Then, for our main result we ignore entrainment, which simplifies the calculation significantly, see 
Sec.\ \ref{sec:trigger}. In Sec.\ \ref{sec:rmode}, where we discuss the analogue of the 
$r$-mode instability, entrainment is included and turns out to be crucial.

In the absence of dissipation, the two-fluid stress-energy tensor is 
\be
T^{\mu\nu}_{\rm ideal} = j_1^\mu p_1^\nu +j_2^\mu p_2^\nu -g^{\mu\nu} \Psi \, . 
\ee
The continuity equations for the two conserved currents are 
\be \label{djmu2}
0 = \partial_\mu j_1^\mu = \partial_\mu j^\mu_2  \, ,
\ee
and we can write the divergence of the (ideal) stress-energy tensor 
as 
\be \label{dTmunu2}
\partial_\mu T^{\mu\nu}_{\rm ideal} = j_{1,\mu} \omega^{\mu\nu}_1 + j_{2,\mu} \omega^{\mu\nu}_2 \, , 
\ee
with the vorticities 
\be \label{vort}
\omega^{\mu\nu}_n = \partial^\mu p_n^\nu - \partial^\nu p_n^\mu \, .
\ee
To derive Eq.\ (\ref{dTmunu2}) we have used the continuity equations (\ref{djmu2}) 
and $d\Psi = j_1^\mu dp_{1,\mu}+j_2^\mu dp_{2,\mu}$. We see that from the total energy-momentum conservation alone only the {\it sum} of the vorticity terms $j_{1,\mu} \omega^{\mu\nu}_1$ and $j_{2,\mu} \omega^{\mu\nu}_2$ (plus dissipative terms which are not included for now) is required to vanish. It is usually argued, for instance starting from a variational principle \cite{Andersson:2006nr}, that instead there are two separate vorticity equations, 
$j_{1,\mu} \omega^{\mu\nu}_1=j_{2,\mu} \omega^{\mu\nu}_2=0$. This corresponds 
to using two separate Euler equations in a non-relativistic two-fluid formalism \cite{1992ApJ...395..250G,Glampedakis:2010ec}. 
If one of the fluids, say fluid 2, is a superfluid, i.e., $p_2^\mu=\partial^\mu\psi$, then its
vorticity vanishes, $\omega^{\mu\nu}_2=0$. In this case, 
the energy-momentum conservation becomes equivalent to the vorticity equation $j_{1,\mu} \omega^{\mu\nu}_1=0$ if fluid 1 is non-dissipative or $j_{1,\mu} \omega^{\mu\nu}_1+\partial_\mu T^{\mu\nu}_{\rm diss}=0$ 
in the presence of dissipation. In Sec.\ \ref{sec:trigger} this is the 
situation we shall consider, i.e., we shall assume fluid 1 to be dissipative and fluid 2 to be a superfluid. 
The main results of that section can also be obtained by simply considering two separate vorticity equations (one of them supplemented by the dissipative terms). The main result of Sec.\ \ref{sec:rmode}, however, relies on using only the {\it total} energy-momentum conservation and not assuming the non-dissipative fluid to be a superfluid. We shall comment on this point in more detail in Sec.\ \ref{sec:rmode}.

For the explicit evaluation of the current conservation equations (\ref{djmu2}) 
we need the derivatives of the coefficients ${\cal B}_1$, ${\cal B}_2$, ${\cal A}$, i.e., the second derivatives of the generalized pressure, which we abbreviate by
\bea \label{baad}
b_n &\equiv& 4\frac{\partial^2 \Psi}{\partial (p_n^2)^2}\, , \qquad a_n \equiv 2\frac{\partial^2 \Psi}{\partial p_n^2\partial p_{12}^2} \, , 
\qquad a_{12} \equiv \frac{\partial^2 \Psi}{\partial (p_{12}^2)^2} \, , \qquad d\equiv 4\frac{\partial^2 \Psi}{\partial p_1^2\partial p_2^2} \, .
\eea
The coefficients $b_1$ and $b_2$ are nonzero even in the absence of a coupling between the fluids and are related to the sound speeds of the fluids. The other coefficients characterize the coupling between the fluids
with $a_1$, $a_2$, $a_{12}$ being nonzero only in the presence of entrainment, while $d$ describes the non-entrainment coupling. 

As in the case of a single fluid,
we introduce fluctuations and linearize the conservation equations.
As our independent fluctuations we choose the chemical potentials 
and velocities in the laboratory frame, $\delta \vec{v}_1$, $\delta\vec{v}_2$, $\delta\mu_1$, $\delta\mu_2$. 
Then, the divergence of the currents becomes 
[we omit the exponential $e^{i(\omega t-\vec{k}\cdot\vec{x})}$] 
\begingroup \allowdisplaybreaks
\bea 
\partial_\mu j_1^\mu &=& \Big[{\cal B}_1 g_{\mu\nu}+b_1 p_{1\mu}p_{1\nu}+a_1(p_{1\mu} p_{2\nu}+p_{2\mu}p_{1\nu})+a_{12} p_{2\mu}p_{2\nu}\Big]\partial^\mu p_1^\nu \non[2ex]
&&+\Big[{\cal A} g_{\mu\nu}+a_1p_{1\mu}p_{1\nu}+d\,p_{1\mu} p_{2\nu}+a_{12}p_{2\mu}p_{1\nu}+a_2p_{2\mu}p_{2\nu}\Big]\partial^\mu p_2^\nu \non[2ex]
&=&i\Big\{\omega_{v_1}{\cal B}_1+\omega_{v_1}\mu_1[\mu_1b_1(1-v_1^2)+\mu_2a_1(1-\vec{v}_1\cdot\vec{v}_2)]+
\omega_{v_2}\mu_2[\mu_1a_1(1-v_1^2)+\mu_2a_{12}(1-\vec{v}_1\cdot\vec{v}_2)]\Big\}\delta\mu_1 \non[2ex]
&&+i\Big\{\omega_{v_2}{\cal A}+\omega_{v_1}\mu_1[\mu_1a_1(1-\vec{v}_1\cdot\vec{v}_2)+\mu_2 d(1-v_2^2)]+
\omega_{v_2}\mu_2[\mu_1a_{12}(1-\vec{v}_1\cdot\vec{v}_2)+\mu_2a_{2}(1-v_2^2)]\Big\}\delta\mu_2\non[2ex]
&&-i(\omega_{v_1}\mu_1b_1+\omega_{v_2}\mu_2a_1)\mu_1^2\vec{v}_1\cdot\delta\vec{v}_1-i(\omega_{v_1}\mu_1d +\omega_{v_2}\mu_2a_2)\mu_2^2\vec{v}_2\cdot\delta\vec{v}_2
-i{\cal B}_1\mu_1\vec{k}\cdot\delta\vec{v}_1-i{\cal A}\mu_2\vec{k}\cdot\delta\vec{v}_2\non[2ex]
&&-i\mu_1\mu_2(\omega_{v_1}\mu_1a_1+\omega_{v_2}\mu_2a_{12})(\vec{v}_1\cdot\delta\vec{v}_2+\vec{v}_2\cdot\delta\vec{v}_1) \, ,
\eea
\endgroup
and the same with $1\leftrightarrow 2$ for $\partial_\mu j_2^\mu$. 
For the energy-momentum conservation we need the temporal and spatial components
\begin{subequations} \allowdisplaybreaks \label{Tmunu120i}
\bea
\partial_\mu T^{\mu 0} &=& i\Big[\mu_1{\cal B}_1(\omega v_1^2-\vec{k}\cdot \vec{v}_1)+\mu_2{\cal A}(\omega\vec{v}_1\cdot\vec{v}_2-\vec{k}\cdot\vec{v}_2)\Big]\delta\mu_1 
 \non[2ex]
&&+i\mu_1(\mu_1{\cal B}_1\omega\vec{v}_1+\mu_2{\cal A}\omega\vec{v}_2)\cdot\delta\vec{v}_1 +(1\leftrightarrow 2) +\partial_\mu T^{\mu 0}_{\rm diss} \\[2ex]
\partial_\mu T^{\mu i} &=& i\Big\{(\mu_1{\cal B}_1\omega_{v_1}+\mu_2{\cal A}\omega_{v_2})v_{1i}-[\mu_1{\cal B}_1(1-v_1^2)+\mu_2{\cal A}(1-\vec{v}_1\cdot\vec{v}_2)]k_i
\Big\}\delta\mu_1 \non[2ex]
&&+i\mu_1(\mu_1{\cal B}_1\omega_{v_1}+\mu_2{\cal A}\omega_{v_2})\delta v_{1i}
+i\mu_1k_i(\mu_1{\cal B}_1\vec{v}_1+\mu_2{\cal A}\vec{v}_2)\cdot\delta\vec{v}_1 +(1\leftrightarrow 2) +\partial_\mu T^{\mu i}_{\rm diss}\, ,
\eea
\end{subequations}
where $(1\leftrightarrow 2)$ stands for repeating all preceding terms with subscripts 1 and 2 exchanged. 
In the non-dissipative part one can recover the two contributions from the
vorticities (\ref{dTmunu2}): all terms containing fluctuations of fluid 1, i.e., 
$\delta\mu_1$ and $\delta\vec{v}_1$, combine to $j_{1,\mu} \omega^{\mu\nu}_1$, while the terms containing fluctuations of fluid 2 combine to the vorticity terms of fluid 2. 

Dissipation is added as explained in the introduction: we use the single-fluid 
dissipative terms for one of the fluids and leave the second fluid unchanged. In particular, we only consider the dissipative coefficients that already appear in single-fluid first-order hydrodynamics, and thus ignore any additional dissipative 
coefficients that occur due to the relative motion of the two fluids. 
Since in this section we have restricted ourselves to 
zero temperature, the
only relevant transport coefficients are $\eta$ and $\zeta$. Choosing without 
loss of generality fluid 1 to be dissipative, the zero-temperature limit of 
of Eqs.\ (\ref{DissExpl}) yields the dissipative contribution,
\begin{subequations} 
\bea
\partial_\mu T^{\mu 0}_{\rm diss} &\simeq& 
 -\gamma_1^3\left[\eta(\omega^2-\gamma_1^2\omega_{v_1}^2-k^2)+\left(\zeta+\frac{\eta}{3}\right)\omega_{v_1}(\omega-\gamma_1^2\omega_{v_1})\right] \vec{v}_1\cdot\delta\vec{v}_1 \non[2ex] &&+\gamma_1\left[\left(\zeta+\frac{\eta}{3}\right)(\omega-\gamma_1^2\omega_{v_1})\right]\vec{k}\cdot\delta\vec{v}_1 \, , \\[2ex]
\partial_\mu T^{\mu i}_{\rm diss} &\simeq& 
-\gamma_1^3\left[\eta v_{1i} (\omega^2-\gamma_1^2\omega_{v_1}^2-k^2)+\left(\zeta+\frac{\eta}{3}\right)\omega_{v_1} (k_i-\gamma_1^2\omega_{v_1} v_{1i})\right] \vec{v}_1\cdot\delta\vec{v}_1 \non[2ex]
&&+\gamma_1\left(\zeta+\frac{\eta}{3}\right)(k_i-\gamma_1^2\omega_{v_1} v_{1i}) \vec{k}\cdot\delta\vec{v}_1 -\gamma_1\eta(\omega^2-\gamma_1^2\omega_{v_1}^2-k^2)\delta v_{1i} \,.
 \eea
\end{subequations}

\subsection{Dynamical instability triggered by dissipation}
\label{sec:trigger}

We  perform the calculation of the sound modes in the rest frame of the 
dissipative fluid, i.e., $\vec{v}_1=0$ (of course, the fluctuations $\delta\vec{v}_1$ 
are allowed to be nonzero). 
In this frame, there are two particular directions: 
the direction of propagation 
of the sound wave $\vec{k}$ and the direction of the (uniform) velocity of the 
ideal fluid $\vec{v}_2$. As in the case of a single fluid we choose the $y$-$z$-plane
such that both $\vec{k}$ and $\vec{v}_2$ lie in this plane, and rather than working with the $x$, $y$, and $z$ components we work with the $x$ component and the two contractions with $\vec{k}$ and $\vec{v}_2$. Therefore, we have 8 scalar fluctuations, 
\be \label{8var}
\delta\mu_1 \, , \quad \delta\mu_2\, , \quad\vec{k}\cdot\delta\vec{v}_1\, , \quad\vec{k}\cdot\delta\vec{v}_2\, , \quad\vec{v}_2\cdot\delta\vec{v}_1\, , \quad\vec{v}_2\cdot\delta\vec{v}_2\, , \quad\delta v_{1x}\, , \quad\delta v_{2x} \, .
\ee
As mentioned above, let us assume 
the non-dissipative fluid to be a superfluid, such that $p_2^\mu =(\partial_t\psi,-\nabla\psi)$. Hence, $\partial_t\vec{p}_2=-\nabla\mu_2$, which, within our plane wave ansatz, results in the constraint for the fluctuations $\omega\delta\vec{p}_2 = \vec{k}\delta\mu_2$ or, equivalently, $\omega\mu_2\delta\vec{v}_2 = (\vec{k}-\omega\vec{v}_2)\delta\mu_2$.
As a consequence, superfluids 
only allow for longitudinal modes, where $\vec{p}$ oscillates in the direction of $\vec{k}$. We can thus express the fluctuations  $\delta\vec{v}_2$ in terms of $\delta\mu_2$, which reduces the number of independent variables to 5, namely $\delta\mu_1$, $\delta\mu_2$, $\vec{k}\cdot\delta\vec{v}_1$, $\vec{v}_2\cdot\delta\vec{v}_1$, $\delta v_{1x}$. 
We find that in this setup the temporal component of the energy-momentum component 
is automatically fulfilled, $\partial_\mu T^{\mu 0}=0$. This leaves us with the 5 
scalar equations $\partial_\mu j_1^\mu=\partial_\mu j_2^\mu=v_{2i} \partial_\mu T^{\mu i}= k_i \partial_\mu T^{\mu i}=\partial_\mu T^{\mu x}=0$, matching the number of variables. We neglect entrainment for now, $a_1=a_2=a_{12}={\cal A}=0$.
As for a single fluid, we find that the equation $\partial_\mu T^{\mu x}=0$ decouples since it is the only equation that contains $\delta v_{1x}$. It yields the mode
$T^+$ for the dissipative fluid, in which the only nonzero oscillation is $\delta v_{1x}$. Since this mode is completely unaffected by the second fluid, it is not relevant for the following discussion. Notice that the unphysical mode 
$T^-$ does not appear because we work in the rest frame of the dissipative fluid.
We are left with 4 equations for 4 variables, which read 
\be
\left(\begin{array}{cccc} \frac{\omega}{p_1c_1^2} & \frac{\Delta_1\gamma_2\omega_{v_2}}{p_2}& -1 & 0 \\[2ex]
\frac{\Delta_2\gamma_2\omega_{v_2}}{p_1} & \frac{c_2^2(\omega^2-k^2)+(1-c_2^2)\gamma_2^2\omega_{v_2}^2}{\omega p_2c_2^2} & 0 & 0 \\[2ex]
\frac{k^2}{p_1} & 0 & -\omega+2ik^2\Gamma_0 & 0 \\[2ex]
\frac{\vec{v}\cdot\vec{k}}{p_1} & 0 & i\vec{v}\cdot\vec{k}\Gamma_2 & -\omega + 2ik^2\Gamma_1 
\end{array}\right)\left(\begin{array}{c} \delta\mu_1 \\[2ex]
\delta \mu_2 \\[2ex] \vec{k}\cdot\delta\vec{v}_1 \\[2ex] \vec{v}_2\cdot\delta\vec{v}_1 \end{array}\right) = 0  \, . \label{4eqs}
\ee
Here we have eliminated ${\cal B}_n$ and $b_n$ in favor of more physical quantities: First, we can relate ${\cal B}_n$ to the densities $n_n=j_n\cdot v_n$ 
in the respective fluid rest frames. With Eqs.\ (\ref{vp}) and (\ref{j1j2})
 this yields, in the absence of entrainment\footnote{With entrainment
 we would have
 \be
 n_1={\cal B}_1p_1+{\cal A}p_2v_1\cdot v_2 = {\cal B}_1p_1+{\cal A}p_1p_2^2\frac{1-\vec{v}_1\cdot\vec{v}_2}{\sqrt{1-v_1^2}\sqrt{1-v_2^2}}
 = {\cal B}_1p_1+{\cal A}p_1p_2^2\left[1+\frac{(\vec{v}_1-\vec{v}_2)^2}{2}+\ldots\right]
 \, , \nonumber
 \ee
 and the same with $1\leftrightarrow 2$ for $n_2$. 
},
\be
{\cal B}_n = \frac{n_n}{p_n} \, . 
\ee
Second, with the sound speed of a single fluid (\ref{csquared}) and
the definition of $b_n$ in Eq.\ (\ref{baad}) we obtain 
\be \label{bc}
c_n^2 = \frac{n_n}{p_n}\left(\frac{\partial n_n}{\partial p_n}\right)^{-1} 
= \frac{{\cal B}_n}{p_n^2b_n+{\cal B}_n} \, .
\ee
In the absence of a coupling between the fluids, $c_1$ and $c_2$ are the sound speeds of the two fluids, measured in their respective rest frames. The coupling 
leads to a different (much more complicated) form of the sound speeds, but also affects the thermodynamic quantities ${\cal B}_n$ and $b_n$ themselves, i.e., even the quantities $c_1$ and $c_2$ as defined in Eq.\ (\ref{bc}) 
change as the coupling is varied. Third, we have introduced the dimensionless "mixed susceptibilities" 
\be
\Delta_{1} \equiv \frac{p_{2}}{n_{1}}\frac{\partial n_{1}}{\partial p_{2}} = \frac{p_{2}^2d}{{\cal B}_{1}} \, ,
\ee
and the same for $1\leftrightarrow 2$,
which are only nonzero in the presence of a coupling between the fluids. 
The attenuation constant $\Gamma_0$ is as defined in Eq.\ (\ref{Gamma0}), and additionally we have abbreviated 
\be
\Gamma_1 \equiv \frac{\eta}{2p_1n_1} \, , \qquad 
\Gamma_2 \equiv \frac{\eta + 3\zeta}{6p_1n_1} \, .
\ee
The system of equations (\ref{4eqs}) is further simplified by the observation that $\vec{v}_2\cdot\delta\vec{v}_1$ only appears in the last equation (i.e., the 4th column of the $4\times 4$ matrix has a nonzero entry only in its last component). This equation can thus be ignored for the calculation of the dispersions. It can later be used 
to compute $\vec{v}_2\cdot\delta\vec{v}_1$, if needed. 
We are left with the determinant of a $3\times 3$ matrix to determine the dispersion relations $\omega(\vec{k}) $. This determinant is a quartic polynomial in $\omega$, yielding 4 modes. In the absence of a coupling between the fluids, these modes simply correspond to the two modes $L^\pm$ for each fluid. When the fluids are coupled, 
the dispersions are very complicated in general. Nevertheless, our main result 
can be extracted from the general result and stated in a very compact form. To this end, as in Sec.\ \ref{sec:single}, we expand the modes in the momentum up to second order, $\omega \simeq ck + i\Gamma k^2$. 
We expect that one of the modes flips over at a certain relative velocity between the fluids, say $v_2=v_{20}$ such that $c(v_{20})=0$. This is qualitatively the same effect as for a boosted single fluid, see Fig.\ \ref{fig:singleUpDown}, and constitutes an energetic instability. However, due to the presence of the second fluid we now find 
that at exactly this point the attenuation constant of this mode changes its sign too, and as a consequence the mode changes from being damped to being {\it dynamically} unstable. This observation can be 
formulated analytically in the vicinity of the critical velocity $v_{20}$.  To linear order, the sound speed close to $v_{20}$ is 
\be \label{ccrit}
c \simeq -\beta\left(1-\frac{v_2}{v_{20}}\right) \, , 
\ee
with 
\be
v_{20} = \frac{c_2}{\sqrt{\cos^2\theta+c_2^2(1-\cos^2\theta)}} \, , \qquad 
\beta = \frac{c_2^2}{v_{20}(1-c_2^2)\cos\theta} \, ,
\ee
where $\theta$ is the angle between the direction of propagation $\vec{k}$ and the 
direction of flow of the non-dissipative fluid $\vec{v}_2$. The  corresponding attenuation is to linear order
\be \label{Gamunstable}
\Gamma \simeq \Gamma_0 \frac{\Delta_1\Delta_2 c_2^4}{(1-c_2^2)}\left(1-\frac{v_2}{v_{20}}\right) \, .
\ee
In particular, close to the critical velocity $\Gamma$ does not depend on the angle $\theta$. 
The dynamical instability sets in first (i.e., at the smallest relative velocity) 
for modes (anti-)parallel to $\vec{v}_2$. In this case, $\cos^2\theta=1$ and thus $v_{20}=c_2$, and
we see that $c$ changes its sign from negative (upstream) to positive (downstream)
as we move through the critical point by increasing the relative velocity, while 
$\Gamma$ changes its sign from positive (damped) to negative (unstable). 
The time scale for the exponential growth of this mode is given by Eq.\ (\ref{Gamunstable}). It depends on the bulk and shear viscosity coefficients, but also on the thermodynamic properties of the two fluids and their coupling strength. We recall that for this result entrainment plays no role. Of course, nonlinear effects will set in as
the amplitude of the mode grows, in the current linear approach we can make no prediction about the fate of the system in the dynamically unstable regime. 

To further illustrate this result we plot all four modes and their attenuation as a 
function of the relative velocity. For this numerical evaluation we need to choose numerical parameters for the quantities appearing in Eq.\ (\ref{4eqs}). If we were to choose random numbers we would have no control over whether the two fluids can coexist, and thermodynamic instabilities -- which we are not interested in -- might manifest themselves in unstable hydrodynamics modes and thus obscure the interpretation of our results. Therefore, we employ the following form of the generalized pressure, 
\be \label{Psi1}
\Psi = \frac{\lambda_1(p_2^2-m_2^2)^2+\lambda_2(p_1^2-m_1^2)^2+2(h+g p_{12}^2)(p_1^2-m_1^2)(p_2^2-m_2^2)}{4[\lambda_1\lambda_2-(h+g p_{12}^2)^2]} \, ,
\ee
which is based on a relativistic field-theoretical model for two complex scalar fields
with masses $m_1$, $m_2$, self-coupling constants $\lambda_1>0$, $\lambda_2>0$, a
non-entrainment coupling $h$ and an entrainment coupling $g$ \cite{Haber:2015exa}.
The pressure (\ref{Psi1}) is valid in the parameter space where 
condensates of both fields coexist, see Ref.\ \cite{Haber:2015exa} 
for a systematic discussion of the parameter space. 
In all following results, we choose our parameters such that at vanishing relative velocity this phase of miscible fluids is stable. Strictly speaking, the pressure (\ref{Psi1}) refers to 
a system of two superfluids because  the vacuum expectation values (= 
Bose-Einstein condensates) of the two scalar fields constitute the fluids. However, for our purpose it is sufficient to consider $\Psi$ as an example for a microscopic equation of state even though we assume one of the fluids to be dissipative. We do not need to provide any microscopic form or specific numbers for the viscosity coefficients because the attenuation $\Gamma$ scales with $\Gamma_0$ (\ref{Gamma0}) and otherwise does not depend on $\eta$ and $\zeta$. 

\begin{figure} [t]
\begin{center}
\hbox{\includegraphics[width=0.5\textwidth]{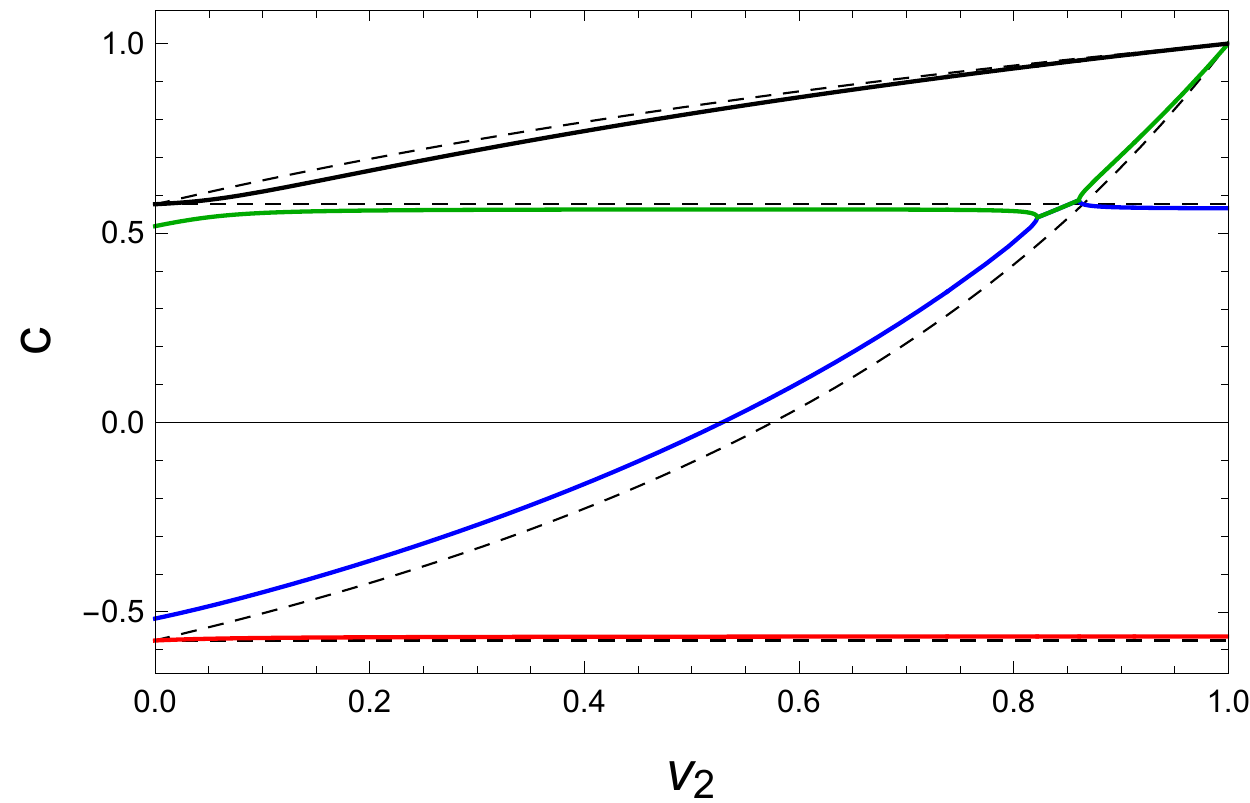}\includegraphics[width=0.5\textwidth]{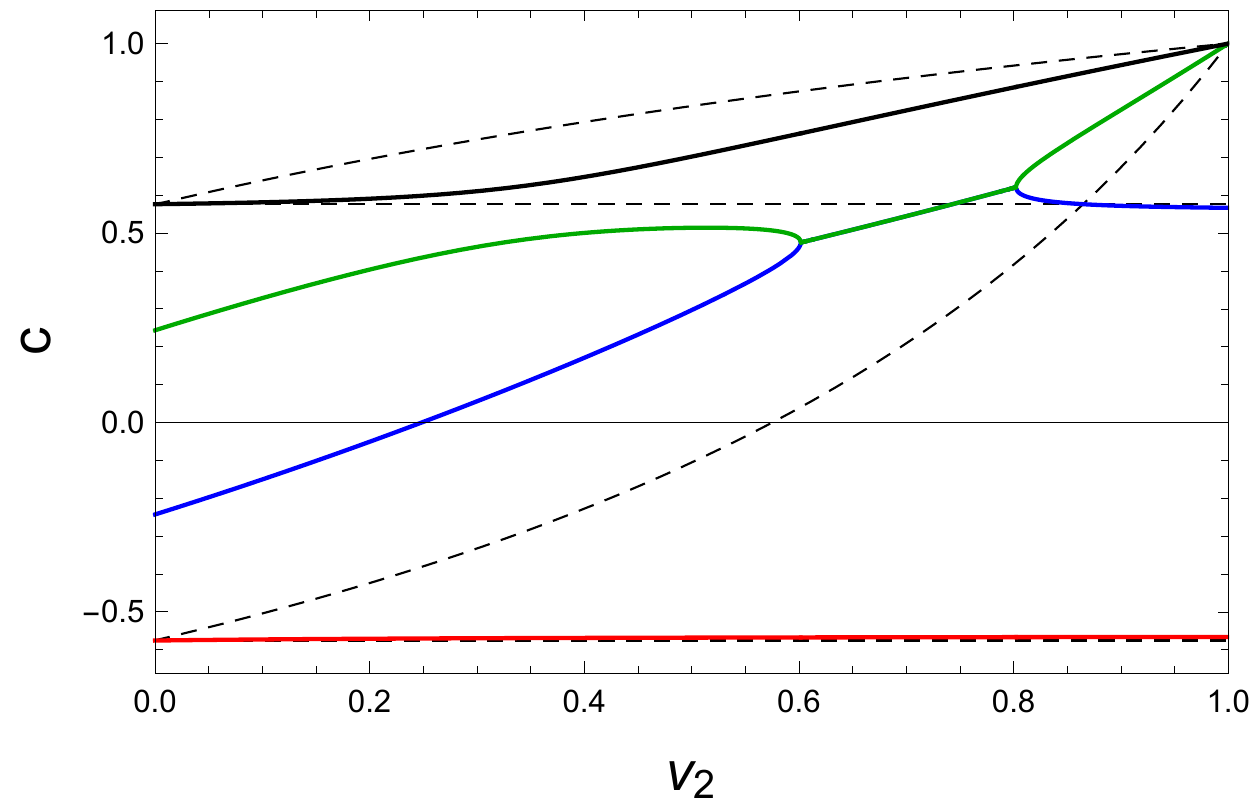}}

\hbox{\includegraphics[width=0.5\textwidth]{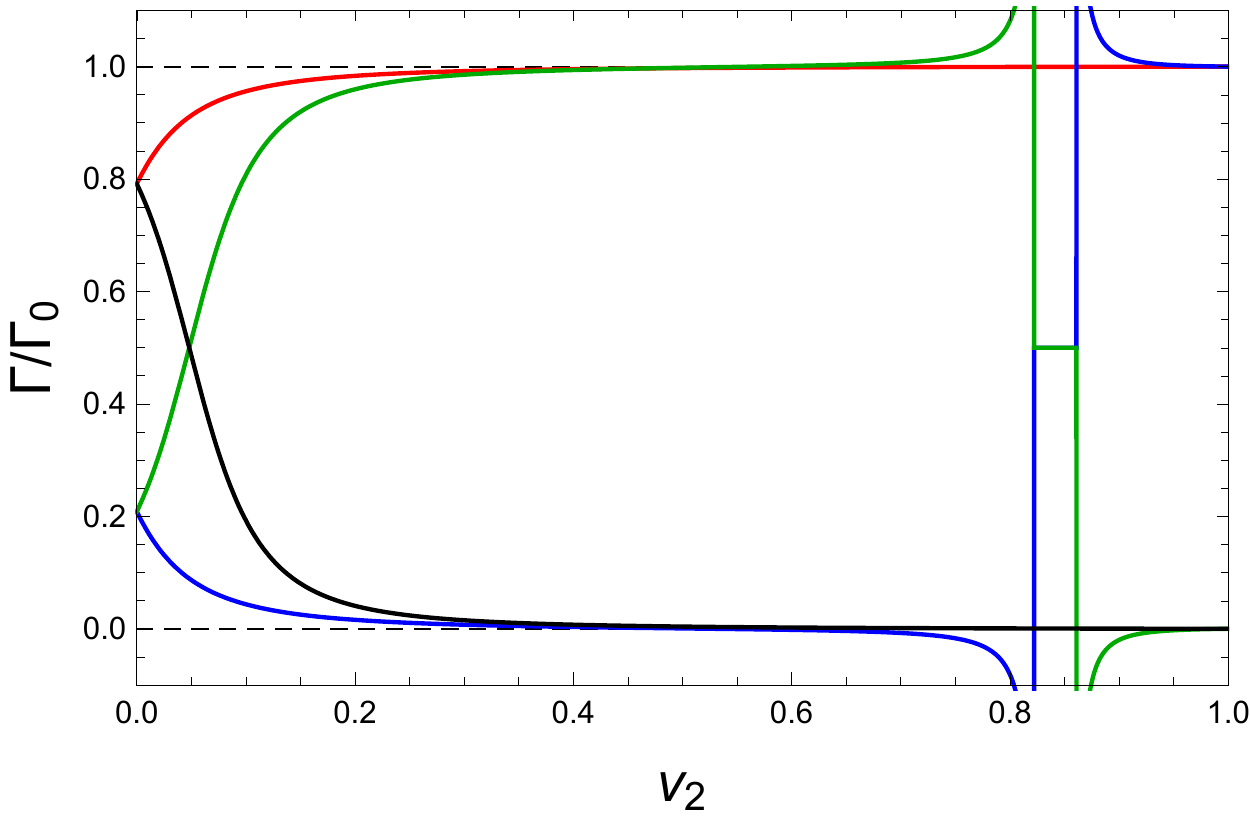}\includegraphics[width=0.5\textwidth]{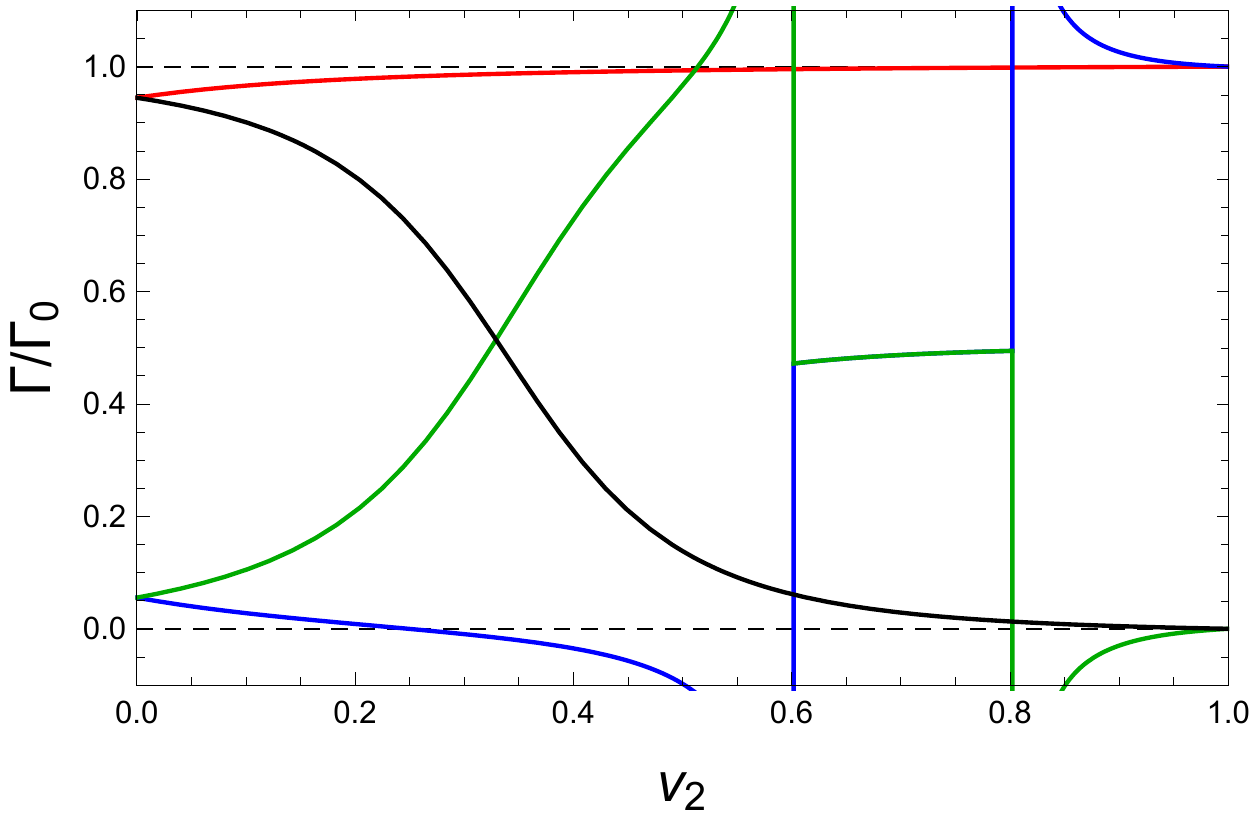}}
\caption{ Real parts of the sound speeds $c$ (upper panels) and corresponding attenuation $\Gamma$ (lower panels, with corresponding colors) for upstream ($c<0$) and downstream ($c>0$) modes as a function of the velocity of the non-dissipative fluid $v_2$ in the rest frame of the dissipative fluid, for weakly coupled fluids (left panels) and strongly coupled fluids (right panels). See text for
the choice of parameters. Thin dashed lines are the modes in the uncoupled system. 
The main result is the change of sign of $\Gamma$ and thus the onset of a dynamical instability at the point where one of the modes (blue) flips over from upstream to downstream.  }
\label{fig:2fluids2}
\end{center}
\end{figure}

We show the result for the 4 upstream/downstream modes in Fig.\ \ref{fig:2fluids2} with parameters $m_1=m_2\equiv m$, $p_1=12m$, $p_2=10m$,  $\lambda_1=0.2$, 
$\lambda_2=0.3$. Since we have ignored entrainment in the hydrodynamic calculation, we obviously have to set the microscopic entrainment coupling to zero, $g=0$, for consistency. The results are shown for weak coupling, $h=-0.03$ (left panels) and strong coupling $h=-0.12$ (right panels). Other than 
creating a stable system of coexisting fluids at vanishing counterflow,
this particular choice of parameters has no specific physical motivation. Somewhat different
results are obtained for $h>0$, but the main qualitative conclusions 
are the same and thus we restrict ourselves to $h<0$.   
The main observations are:
\begin{itemize}
\item For both weak and strong coupling the figure shows the onset of the 
dynamical instability at the point where the behavior of $c$ indicates an energetic instability. This confirms the analytical results (\ref{ccrit}) -- (\ref{Gamunstable}).

\item The sound speeds themselves show a dynamical instability at large velocities: where the real parts of two branches coincide, their imaginary parts
(not shown in the figure) have opposite signs, i.e., one of the modes grows exponentially. Since the results linear in $k$ 
are not affected by dissipative effects, this is the same observation as already made in Refs.\ \cite{2004MNRAS.354..101A,Haber:2015exa}. As a consequence, the system is dynamically unstable for all $v_2>v_{20}$: First due to 
${\rm Re}\, \Gamma <0$ with a growth time for the instability $\tau = (k^2|{\rm Re}\, \Gamma|)^{-1}$, i.e., at the onset the growth time is infinite. Then 
due to ${\rm Im}\, c <0$ with growth time $\tau = (k|{\rm Im}\, c|)^{-1}$, while there is no instability in the $k^2$ terms in this regime, i.e., ${\rm Re}\, \Gamma >0$.
Interestingly, just before and just after this regime ${\rm Re}\, \Gamma$ diverges, indicating an extreme instability with infinitesimally short growth time. Then, at the highest velocities, the instability is given again by the second-order contribution, ${\rm Re}\, \Gamma <0$.

\item There is a "role reversal" between damped and undamped modes as a function of the counterflow velocity: For small $v_2$ there is a pair of upstream modes ($c>0$)
and a pair of downstream modes ($c<0$). Each pair has a strongly damped and weakly damped mode. As $v_2$ increases, the strongly damped mode becomes weakly damped and vice versa. For small coupling, this role reversal happens in the stable regime, i.e., before the dynamical instability sets in.

\end{itemize}

\subsection{Analogue of the $r$-mode instability}
\label{sec:rmode}

The dynamical instability just discussed sets in at a 
certain nonzero critical velocity. As explained 
in Sec.\ \ref{sec:intro}, this is analogous to modes in rotating stars, such as $f$-modes, which become unstable at a nonzero angular velocity of the star. In contrast, $r$-modes -- whose restoring force is the Coriolis force -- only exist in a rotating star and
can become unstable at arbitrarily small angular velocities. 
Exploiting this analogy, it seems natural to ask whether there is an unstable mode in the present two-fluid system with the same properties.

As before, we work in the rest frame of the dissipative fluid. The 
conservation equations for the two currents and the 
stress-energy tensor yield the 6 equations $\partial_\mu j_1^\mu=\partial_\mu j_2^\mu=\partial_\mu T^{\mu 0}=v_{2i} \partial_\mu T^{\mu i}= k_i \partial_\mu T^{\mu i}=\partial_\mu T^{\mu x}=0$. Now, in contrast to the previous subsection, we do {\it not} assume the non-dissipative fluid to be a superfluid, i.e., we allow for both longitudinal and transverse oscillations and thus there are a priori no constraints on the 8 variables (\ref{8var}). Neither do we consider 
two separate vorticity equations (one of which would be supplemented by the dissipative terms). This is in accordance with all conservation laws of the system, but  
seems to contradict the usual approach of separating the vorticity 
equations (or Euler equations in the 
non-relativistic limit), as already briefly discussed below  Eq.\ (\ref{vort}). The idea behind separate vorticity equations is that the exchange of energy and momentum between the two fluids occurs on
larger length and time scales than the exchange within the fluids. By using a single vorticity equation we assume that this difference in scales is not too large, but still sufficiently large to justify the two-fluid picture with independent velocity fields. It is an interesting general question whether such a regime exists, and we leave a more detailed study of this question for the future. Here,
our main purpose is to point out a solution of the hydrodynamic equations that shows a dynamical instability for arbitrarily small counterflow. At the very least, this is an interesting toy version of the $r$-mode instability in neutron stars. Additionally, it 
would be very interesting to see whether it can be realized in 
a real-world two-fluid system with counterflow.

\begin{figure} [t]
\centering

\hbox{\includegraphics[width=0.33\textwidth]{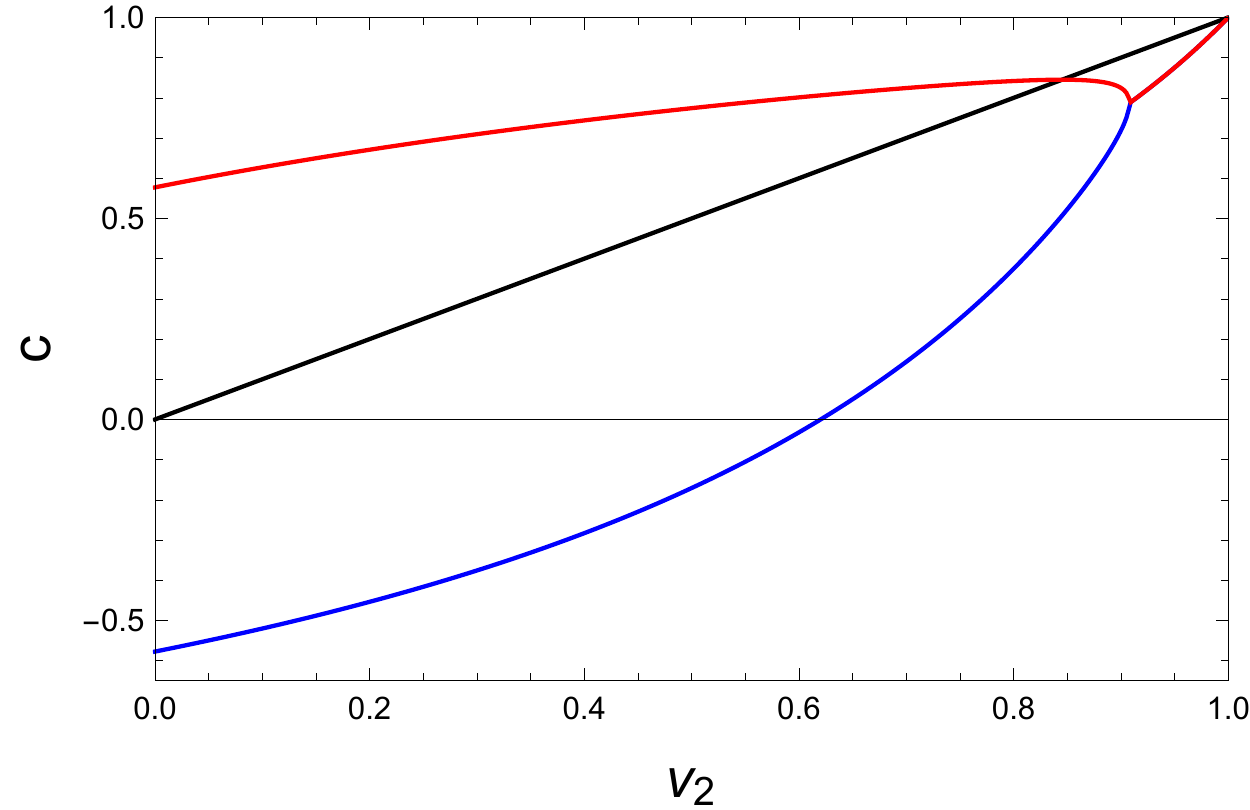}\includegraphics[width=0.33\textwidth]{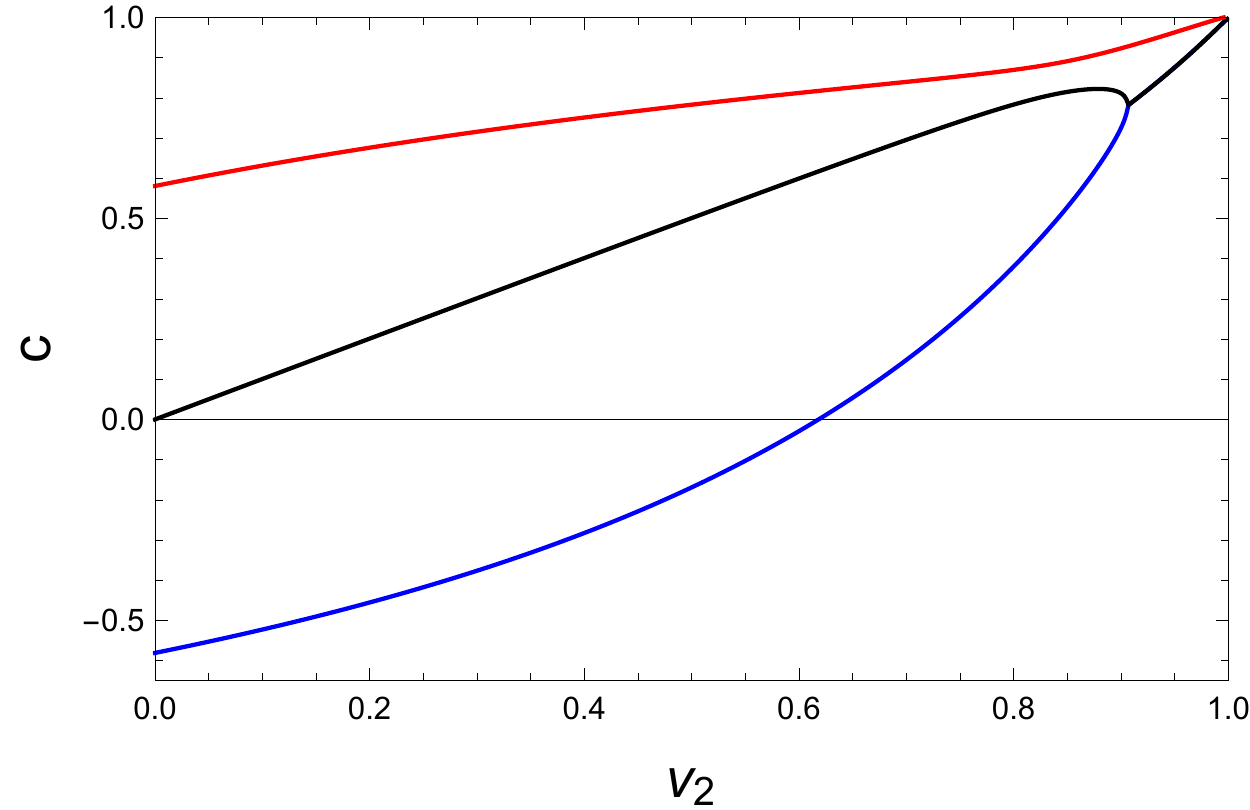}\includegraphics[width=0.33\textwidth]{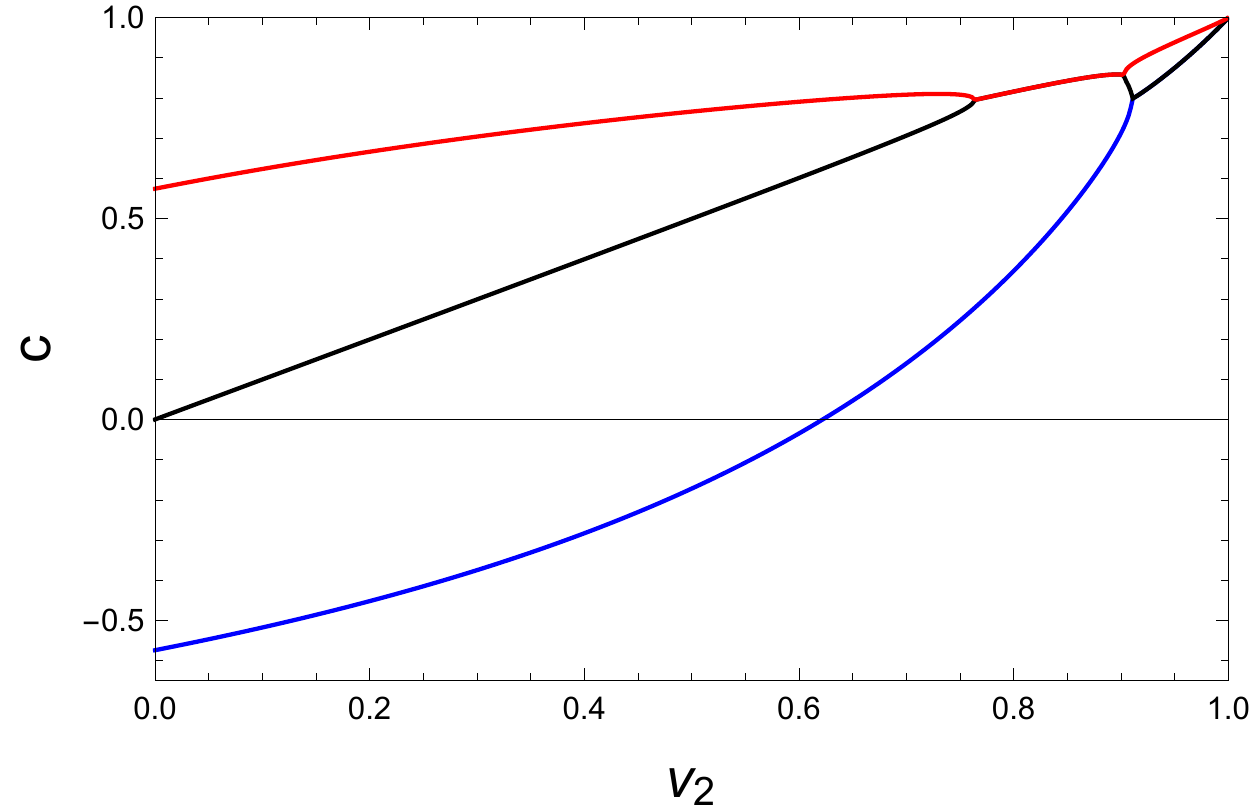}}

\hbox{\includegraphics[width=0.33\textwidth]{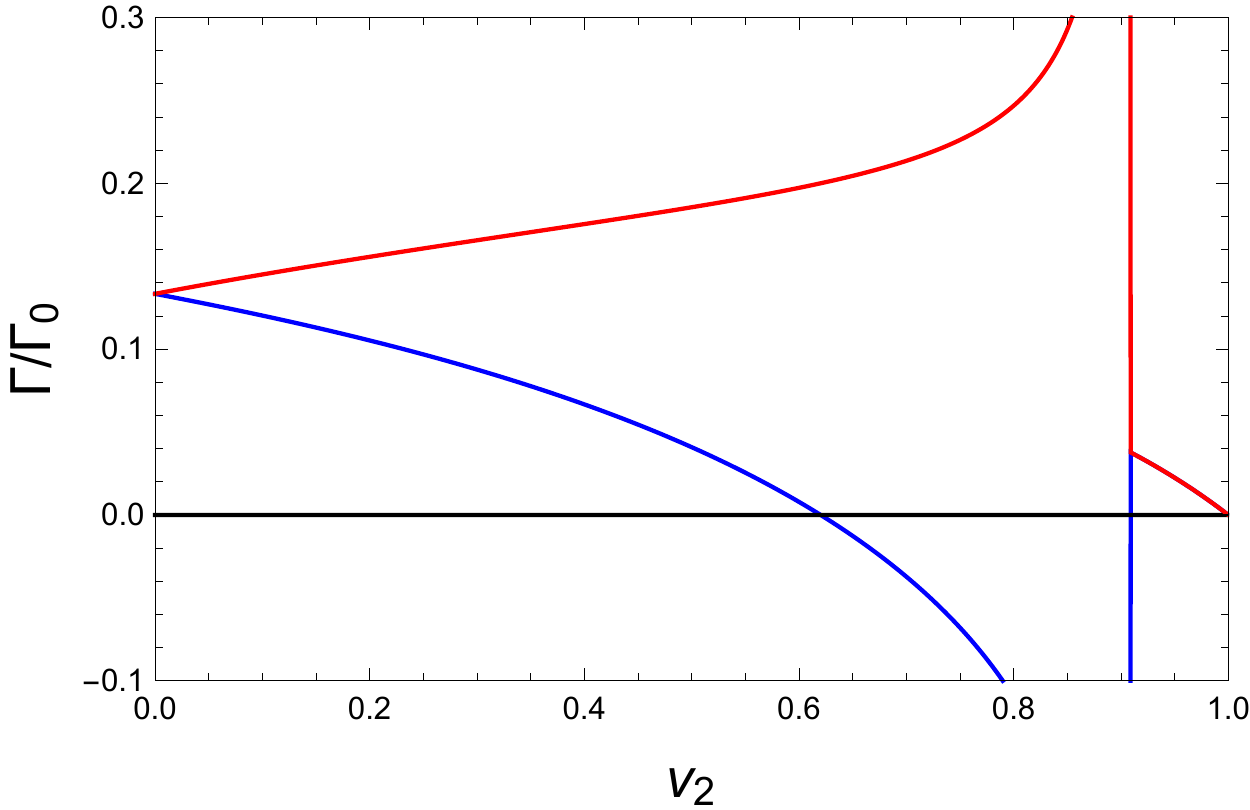}\includegraphics[width=0.33\textwidth]{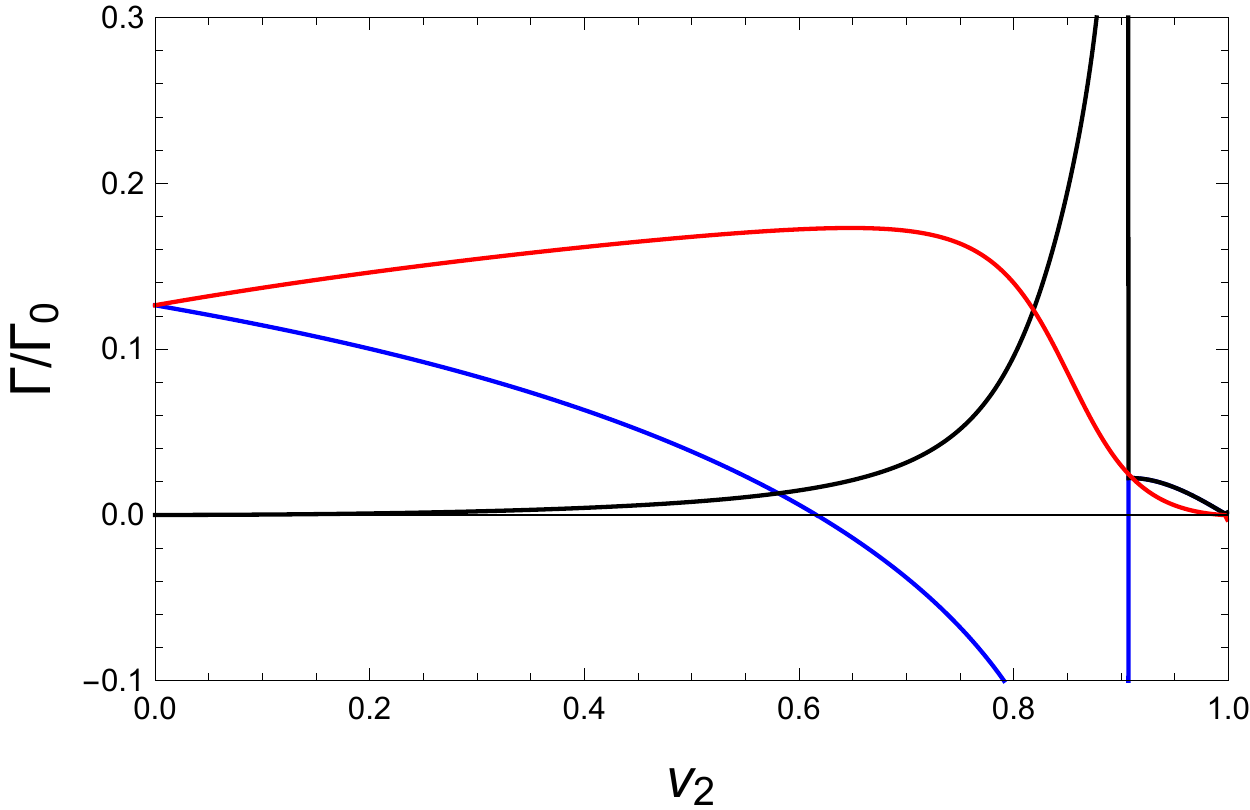}\includegraphics[width=0.33\textwidth]{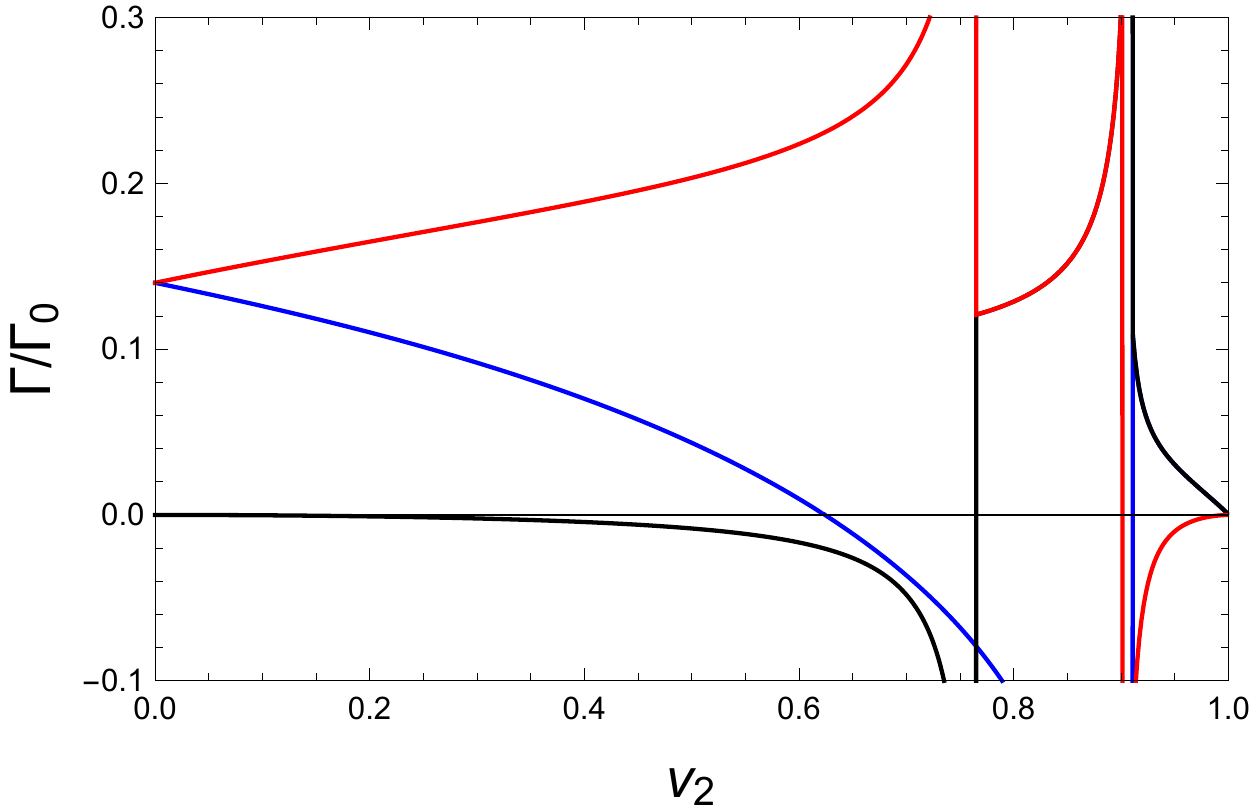}}

\caption{Real parts of speeds (upper panels) and attenuation 
(lower panels) of three modes in a coupled two-fluid system in which 
the chemical potential of the dissipative fluid is clamped, $\delta\mu_1=0$. (There is a fourth, purely diffusive, mode, which is not shown here.) The entrainment coupling between the fluids is zero
(left panels), negative (middle panels), and positive (right panels). 
The (black) mode which only propagates in the presence of a nonzero relative velocity $v_2$ is dynamically unstable ($\Gamma<0$) for arbitrarily small $v_2$ if the entrainment coupling is positive.}
\label{fig:entrain}
\end{figure}

In order to calculate the possible modes of the system, we first note that, as above, the equation $\partial_\mu T^{\mu x}=0$ decouples
because it only contains the fluctuations $\delta v_{1x}$ and $\delta v_{2x}$, and these fluctuations appear nowhere else. We are thus left with 5 equations for 6 fluctuation variables. We find propagating modes if we impose the additional constraint $\delta\mu_1=0$ or $\delta\mu_2=0$. In the following we shall restrict ourselves to the case $\delta\mu_1=0$, i.e., we are looking for modes in which the chemical potential of the dissipative fluid does not oscillate. (The main qualitative result can also be obtained with $\delta\mu_2=0$.) Such constraints 
on the oscillations can in principle be realized experimentally. An 
example is the so-called fourth sound in a superfluid, where 
the velocity of the normal component is clamped by an external force 
and thus cannot oscillate \cite{PhysRev.113.962,khala,Yarom:2009uq}.

We thus consider 5 hydrodynamic 
equations plus the additional equation $\delta\mu_1=0$.
The determinant of the matrix of this system of equations yields  
a fourth-order polynomial from which four modes are calculated. 
For the main observation it is crucial to include entrainment. This renders the 
calculation very tedious, and we have not found any compact analytical results without making use of an explicit equation of state. 
Let us therefore first plot the results using the generalized pressure (\ref{Psi1}) and a particular parameter set, and then show the main result in an analytical form for this specific equation of state. In Fig.\ \ref{fig:entrain}, the upstream/downstream modes are shown for the parameters $p_1=p_2\equiv p$, $m_1=m_2=0$, $\lambda_1=0.2$, $\lambda_2=0.3$, $h=0.05$ and three
different entrainment couplings $g=0$, $g=-0.002/p^2$, $g=+0.002/p^2$.
The parameters are chosen such that at zero counterflow the system of coexisting fluids is stable. Again, we do not need to specify any values for the 
dissipative coefficients since the attenuation of the modes we consider scales with $\Gamma_0$. In the presence of entrainment, $\Gamma_0$ as defined in Eq.\ (\ref{Gamma0}) depends on the relative velocity $v_2$;
for the normalization in Fig.\ \ref{fig:entrain} we use $\Gamma_0(v_2=0)$.  We plot only three of the four modes since the fourth mode is a purely diffusive mode which does not couple to the 
other modes. 

First of all we observe the phenomenon already discussed in Sec.\ \ref{sec:trigger}: there is one mode that becomes dynamically unstable at a nonzero critical velocity. This is independent of entrainment. The new observation is related to the mode that, in the absence of entrainment, is given by $\omega = v_2\cos\theta$. This is the mode $T^+$ in Table \ref{table0}. Since we work in the rest frame of the dissipative fluid, it is the ideal fluid that is in motion and thus we expect this mode to be undamped if there is no coupling to the fluid at rest. This expectation is borne out in the left plots, where the entrainment is set to zero. We see that even in the presence of a non-entrainment coupling $h$, which is nonzero in all plots of the figure,  the mode does not couple to any other modes and thus its 
attenuation vanishes. Only if we switch on an entrainment coupling $g$ this mode couples to the other modes. This mode mixing was already pointed out in Ref.\ \cite{Haber:2015exa}. Here, in the presence of dissipation, we see that this mode acquires a nonzero attenuation if the entrainment coupling is switched on. Depending on the sign of $g$, the mode 
is damped (middle plots, $g<0$) or becomes dynamically unstable (right plots, positive $g>0$). 
For small $g$ [compared to $p_1^{-2}$, $p_2^{-2}$, $(p_1p_2)^{-1}$]
and small $v_2$ and setting $h=m_1=m_2=0$ for simplicity
we find $\omega\simeq ck+i\Gamma k^2$ with 
\be
c\simeq \left(1-\frac{gp_1^3}{2\lambda_1 p_2}\right)v_2\cos\theta
\, , \qquad \Gamma \simeq -\frac{g(4\eta+3\zeta)v_2^2\cos^2\theta}{3p_1 p_2}  \, . 
\ee
This confirms analytically that this mode only propagates in the presence of a relative velocity between the fluids and becomes dynamically unstable 
for $g>0$ and arbitrarily small relative velocities.

\section{Non-relativistic limit}
\label{sec:nonrel}

The purpose of this section is to start from the non-relativistic hydrodynamic equations and re-derive the main result of this paper in this limit. As we have seen in Sec.\ \ref{sec:single}, first-order 
{\it relativistic} hydrodynamics gives rise to unphysical instabilities. 
We have argued that our two-fluid calculation does not suffer from this problem since we have worked in the rest frame of the dissipative fluid and at zero temperature, and thus we expect the dynamical instability we have found to survive in higher-order hydrodynamics.  Nevertheless, it is useful to confirm our result in the non-relativistic case, where there are no unphysical instabilities.
Here we focus on the non-relativistic version of the results of Sec.\ \ref{sec:trigger} by considering two separate Navier-Stokes equations for the two fluids and neglecting entrainment, i.e., we do not set up a non-relativistic version of the analogue of the $r$-mode instability from Sec.\ \ref{sec:rmode}.

\subsection{Single fluid}

As in the relativistic case, let us establish our formalism and notation for a single fluid before moving on to two coupled fluids. 
In the non-relativistic limit, the hydrodynamic equations for a single fluid are 
\cite{Landau:Fluid}
\begin{subequations} \allowdisplaybreaks \label{hydroNR}
\bea
\frac{\partial\rho}{\partial t} + \nabla\cdot\vec{g} &=& 0 \,, \label{contNR}\\[2ex]
\frac{\partial\epsilon}{\partial t} +\nabla\cdot\vec{q} &=& 0 \, , \\[2ex]
\frac{\partial g_i}{\partial t} +\partial_j\Pi_{ji} &=& 0 \, .
\eea
\end{subequations}
(For a derivation of these equations from the relativistic formalism  
see for instance Ref.\ \cite{Schmitt:2014eka}.) 
 Here, $\rho$ is the mass density of the fluid, $\vec{g} = \rho \vec{v}$ is the momentum density with the three-velocity $\vec{v}$, and $\epsilon = \epsilon_0 + \rho v^2/2$ is the energy density with the energy density in the rest frame of the fluid $\epsilon_0$. The energy flux $\vec{q}$ and the stress tensor $\Pi_{ij}$ contain dissipative contributions and are given by 
\be
q_i = (\epsilon+P)v_i +v_j\pi_{ij}+j_{T,i} \, , \qquad \Pi_{ij} = \rho v_i v_j+\delta_{ij}P+\pi_{ij}  \, , 
\ee
with 
\be
\vec{j}_T = -\kappa\nabla T \, , \qquad \pi_{ij} = -\eta\left(\partial_iv_j+\partial_jv_i-\frac{2}{3}\delta_{ij}\nabla\cdot\vec{v}\right)-\zeta\delta_{ij}\nabla\cdot\vec{v} \, . 
\ee
To rewrite Eqs.\ (\ref{hydroNR}) we use the thermodynamic relations $\epsilon_0+P=\mu\rho+sT$ and $d\epsilon_0 = \mu d\rho+Tds$, where $\mu$ is the chemical potential per unit mass. (This definition of $\mu$ is common and useful in the non-relativistic literature, although here it is a slight abuse of notation since in the relativistic calculation we denoted the chemical potential itself by $\mu$.) Then we obtain the equivalent, perhaps more familiar, set of equations in the form of continuity equation, entropy production, and Navier-Stokes equation, 
\begin{subequations}
\bea
\frac{\partial\rho}{\partial t} + \nabla\cdot\vec{g} &=& 0 \,, \\[2ex]
\frac{\partial s}{\partial t} +\nabla\cdot\left(s\vec{v}+\frac{\vec{j}_T}{T}\right) &=& -\frac{\pi_{ji}\partial_j v_i+\vec{j}_T\cdot\nabla T/T}{T} \, , \label{Entropy}\\[2ex]
\frac{\partial v_i}{\partial t}+(\vec{v}\cdot\nabla)v_i&=& - \frac{\partial_i P}{\rho}-\frac{\partial_j\pi_{ji}}{\rho} \, .
\eea
\end{subequations}
Next, we introduce the fluctuations in $\mu$, $T$ and $\vec{v}$ as 
in Eqs.\ (\ref{fluc}) and use $dP=\rho d\mu+sdT$ to find 
\begin{subequations}
\bea
0&=& \omega_v\left(\frac{\partial\rho}{\partial\mu}\delta\mu+\frac{\partial s}{\partial\mu}\delta T \right) - \rho\vec{k}\cdot\delta\vec{v} \, , \label{NR1} \\[2ex]
0&\simeq& \omega_v\left(\frac{\partial\rho}{\partial T}\delta\mu+\frac{\partial s}{\partial T}\delta T \right) -s\vec{k}\cdot\delta\vec{v}-i\frac{\kappa}{T}k^2\delta T \, ,  \label{NR2} \\[2ex]
0&=&-\omega_v\delta v_i +k_i\left(\delta\mu+\frac{s}{\rho}\delta T\right) +i\frac{\eta+3\zeta}{3\rho}k_i\vec{k}\cdot\delta\vec{v}+i\frac{\eta}{\rho}k^2\delta v_i \, ,  \label{NR3} 
\eea
\end{subequations}
where we have neglected the right-hand side of Eq.\ (\ref{Entropy}), which is of second order in the fluctuations.  As explained 
above Eq.\ (\ref{omegai}), we work with the independent velocity  fluctuations $\delta v_x$, $\vec{v}\cdot\delta\vec{v}$, $\vec{k}\cdot\delta\vec{v}$. The transverse component of Eq.\ (\ref{NR3}) decouples from the rest as usual and gives a mode where only
$\delta v_x$ is nonzero with dispersion   
\be \label{mode1NR}
\omega = \vec{v}\cdot\vec{k}+i\frac{\eta}{\rho}k^2 \, .
\ee
The remaining modes are found by Eqs.\ (\ref{NR1}), (\ref{NR2}) together with the two equations obtained from contracting Eq.\ (\ref{NR3}) with $\vec{k}$ and $\vec{v}$,  
\be
\left(\begin{array}{cccc} \frac{\omega_v}{c^2} & \frac{\omega_v}{\rho}\frac{\partial s}{\partial \mu} & -1 & 0 \\[2ex] \omega_v\frac{\partial\rho}{\partial T} & 
\omega_v\frac{\partial s}{\partial T} -ik^2\frac{\kappa}{T} & -s & 0 \\[2ex] k^2 & k^2 \frac{s}{\rho} & -\omega_v+2ik^2\Gamma_0 & 0 \\[2ex]
\vec{v}\cdot\vec{k} & \vec{v}\cdot\vec{k} \frac{s}{\rho} & 2i\Gamma_2\vec{v}\cdot\vec{k} & -\omega_v+2i\Gamma_1k^2 \end{array}\right)\left(\begin{array}{c} \delta\mu \\[2ex]
\delta T \\[2ex] \vec{k}\cdot\delta\vec{v} \\[2ex] \vec{v}\cdot\delta\vec{v} \end{array}\right) = 0  \, ,
\ee
where 
\be
c^2 = \rho\left(\frac{\partial\rho}{\partial \mu}\right)^{-1} 
\ee
is the squared speed of sound in the rest frame of the fluid and, in analogy to the relativistic case, we have defined the attenuation constants 
\be
\Gamma_0 \equiv \frac{4\eta + 3\zeta}{6\rho} \, , \qquad \Gamma_1 \equiv \frac{\eta}{2\rho} \,, \qquad \Gamma_2 \equiv \frac{\eta + 3\zeta}{6\rho} \, .
\ee
The determinant of the $4\times 4$ matrix yields a fourth-order polynomial in $\omega$, such that in total there are 5 modes, all 
of which are stable. 

At $T=0$, the second column
and the second row of the $4\times 4$ matrix vanish, and we find the 3 solutions
\be \label{omnr}
\omega = \pm k\sqrt{c^2-k^2\Gamma_0^2} + \vec{v}\cdot\vec{k} +ik^2\Gamma_0 \, , \qquad \omega = \vec{v}\cdot\vec{k}+i\frac{\eta}{\rho}k^2 \, .
\ee
Together with the mode (\ref{mode1NR}) these are 4 (stable) modes.
They are to be compared to the 4 modes $L^\pm$ and $T^\pm$ in Table \ref{table0},
one of which suffers an unphysical instability. We see that the only effect of the nonzero velocity is to add $ \vec{v}\cdot\vec{k}$ to the dispersion. In particular, the attenuation is independent of $\vec{v}$. We can obtain the modes (\ref{omnr}) from the fully relativistic modes shown in Fig.\ \ref{fig:singleUpDown} by taking the small-velocity limit, i.e., neglecting terms quadratic in the velocity, which includes neglecting terms of order $vc$. [See also Eq.\ (\ref{cGamLor}), which shows that the attenuation becomes independent of the velocity if terms of order $v^2$, $vc$ and higher are neglected.]

\subsection{Two fluids}

For the system of two coupled fluids we restrict ourselves to 
zero temperature and neglect entrainment. 
The pressure is a function of both chemical potentials such that 
$dP = \rho_1 d\mu_1 + \rho_2 d\mu_2$. We consider separate Navier-Stokes equations for the two fluids such that the relevant system of equations is 
\begin{subequations}
\bea
\frac{\partial\rho_1}{\partial t} + \nabla\cdot\vec{g}_1 &=& 0 \,, \\[2ex]
\frac{\partial\rho_2}{\partial t} + \nabla\cdot\vec{g}_2 &=& 0 \,, \\[2ex]
\frac{\partial v_{1i}}{\partial t}+(\vec{v}_1\cdot\nabla)v_{1i}&=& - \partial_i \mu_1-\frac{\partial_j\pi_{ji}}{\rho_1} \,, \\[2ex]
\frac{\partial v_{2i}}{\partial t}+(\vec{v}_2\cdot\nabla)v_{2i}&=& - \partial_i \mu_2 \,,
\eea
\end{subequations}
where, as in the relativistic case, we have chosen fluid 1 to be dissipative, and where $\vec{g}_n = \rho_n\vec{v}_n$. (In the presence of entrainment $\vec{g}_1$ would receive a contribution from $\vec{v}_2$ and vice versa.)  
Introducing the fluctuations as usual, working in the rest frame of the dissipative fluid, $\vec{v}_1=0$, and contracting the Navier-Stokes  equations with $\vec{k}$ and $\vec{v}_2$ 
yields the 6 equations
\be
\left(\begin{array}{cccccc} \frac{\omega}{c_1^2} & \omega\Delta_1 & -1 &0&0&0 \\[2ex]
\omega_{v_2}\Delta_2&  \frac{\omega_{v_2}}{c_2^2} &0&-1&0&0 \\[2ex]
k^2&0&-\omega+2ik^2\Gamma_0 &0&0&0\\[2ex]
0&k^2&0&-\omega_{v_2} &0&0 \\[2ex]
\vec{k}\cdot\vec{v}_2&0&i\Gamma_2\vec{v}_2\cdot\vec{k} &0&-\omega+i\Gamma_1 k^2&0\\[2ex]
0&\vec{k}\cdot\vec{v}_2&0&0&0&-\omega_{v_2} \end{array}\right)\left(\begin{array}{c} \delta\mu_1 \\[2ex] \delta\mu_2\\[2ex]
\vec{k}\cdot\delta\vec{v}_1 \\[2ex] \vec{k}\cdot\delta\vec{v}_2 \\[2ex]\vec{v}_2\cdot\delta\vec{v}_1 \\[2ex]\vec{v}_2\cdot\delta\vec{v}_2\end{array}\right) = 0  \, ,
\ee
where, in analogy to the relativistic calculation, 
we have introduced "mixed susceptibilities"
\be
\Delta_{1} = \frac{1}{\rho_{1}}\frac{\partial \rho_{1}}{\partial \mu_{2}} \, ,
\ee
and the same for $1\leftrightarrow 2$. We find one undamped mode $\omega = \vec{v}_2\cdot\vec{k}$ and one purely diffusive mode $\omega= i\Gamma_1 k^2$. The remaining modes are determined by the $4\times 4$ subdeterminant
in the upper left corner, which yields a polynomial of fourth order in $\omega$. We expand the dispersion as usual for small momenta, $\omega = ck+i\Gamma k^2$, and find that one of these four modes 
flips over (sign change of $c$) at
\be
v_{20} = \frac{c_2}{\cos\theta} \, ,
\ee
while the corresponding attenuation constant behaves in the vicinity of this point as 
\be
\Gamma \simeq \Gamma_0\Delta_1\Delta_2 c_2^4\left(1-\frac{v_2}{v_{20}}\right) \, . 
\ee
Comparing with Eq.\ (\ref{Gamunstable}) we have rederived the 
dynamical instability in the limit of small $c_2$. This confirms 
that the observation that the energetic instability turns dynamical in the presence of dissipation is not an artifact of first-order relativistic hydrodynamics.

\acknowledgments{A.S.\ is supported by the Science \& Technology Facilities Council (STFC) in the form of an Ernest Rutherford Fellowship. N.A. gratefully acknowledges support from STFC through grant number ST/R00045X/1.}

\bibliography{references}

\begin{thebibliography}{58}
\expandafter\ifx\csname natexlab\endcsname\relax\def\natexlab#1{#1}\fi
\expandafter\ifx\csname bibnamefont\endcsname\relax
  \def\bibnamefont#1{#1}\fi
\expandafter\ifx\csname bibfnamefont\endcsname\relax
  \def\bibfnamefont#1{#1}\fi
\expandafter\ifx\csname citenamefont\endcsname\relax
  \def\citenamefont#1{#1}\fi
\expandafter\ifx\csname url\endcsname\relax
  \def\url#1{\texttt{#1}}\fi
\expandafter\ifx\csname urlprefix\endcsname\relax\def\urlprefix{URL }\fi
\providecommand{\bibinfo}[2]{#2}
\providecommand{\eprint}[2][]{\url{#2}}

\bibitem[{\citenamefont{{Ferrier-Barbut}
  et~al.}(2014)\citenamefont{{Ferrier-Barbut}, {Delehaye}, {Laurent}, {Grier},
  {Pierce}, {Rem}, {Chevy}, and {Salomon}}}]{2014Sci...345.1035F}
\bibinfo{author}{\bibfnamefont{I.}~\bibnamefont{{Ferrier-Barbut}}},
  \bibinfo{author}{\bibfnamefont{M.}~\bibnamefont{{Delehaye}}},
  \bibinfo{author}{\bibfnamefont{S.}~\bibnamefont{{Laurent}}},
  \bibinfo{author}{\bibfnamefont{A.~T.} \bibnamefont{{Grier}}},
  \bibinfo{author}{\bibfnamefont{M.}~\bibnamefont{{Pierce}}},
  \bibinfo{author}{\bibfnamefont{B.~S.} \bibnamefont{{Rem}}},
  \bibinfo{author}{\bibfnamefont{F.}~\bibnamefont{{Chevy}}}, \bibnamefont{and}
  \bibinfo{author}{\bibfnamefont{C.}~\bibnamefont{{Salomon}}},
  \bibinfo{journal}{Science} \textbf{\bibinfo{volume}{345}},
  \bibinfo{pages}{1035} (\bibinfo{year}{2014}), \eprint{1404.2548}.

\bibitem[{\citenamefont{{Delehaye} et~al.}(2015)\citenamefont{{Delehaye},
  {Laurent}, {Ferrier-Barbut}, {Jin}, {Chevy}, and
  {Salomon}}}]{2015PhRvL.115z5303D}
\bibinfo{author}{\bibfnamefont{M.}~\bibnamefont{{Delehaye}}},
  \bibinfo{author}{\bibfnamefont{S.}~\bibnamefont{{Laurent}}},
  \bibinfo{author}{\bibfnamefont{I.}~\bibnamefont{{Ferrier-Barbut}}},
  \bibinfo{author}{\bibfnamefont{S.}~\bibnamefont{{Jin}}},
  \bibinfo{author}{\bibfnamefont{F.}~\bibnamefont{{Chevy}}}, \bibnamefont{and}
  \bibinfo{author}{\bibfnamefont{C.}~\bibnamefont{{Salomon}}},
  \bibinfo{journal}{Phys. Rev. Lett.} \textbf{\bibinfo{volume}{115}},
  \bibinfo{eid}{265303} (\bibinfo{year}{2015}), \eprint{1510.06709}.

\bibitem[{\citenamefont{Yao et~al.}(2016)\citenamefont{Yao, Chen, Wu, Liu,
  Wang, Jiang, Deng, Chen, and Pan}}]{PhysRevLett.117.145301}
\bibinfo{author}{\bibfnamefont{X.-C.} \bibnamefont{Yao}},
  \bibinfo{author}{\bibfnamefont{H.-Z.} \bibnamefont{Chen}},
  \bibinfo{author}{\bibfnamefont{Y.-P.} \bibnamefont{Wu}},
  \bibinfo{author}{\bibfnamefont{X.-P.} \bibnamefont{Liu}},
  \bibinfo{author}{\bibfnamefont{X.-Q.} \bibnamefont{Wang}},
  \bibinfo{author}{\bibfnamefont{X.}~\bibnamefont{Jiang}},
  \bibinfo{author}{\bibfnamefont{Y.}~\bibnamefont{Deng}},
  \bibinfo{author}{\bibfnamefont{Y.-A.} \bibnamefont{Chen}}, \bibnamefont{and}
  \bibinfo{author}{\bibfnamefont{J.-W.} \bibnamefont{Pan}},
  \bibinfo{journal}{Phys. Rev. Lett.} \textbf{\bibinfo{volume}{117}},
  \bibinfo{pages}{145301} (\bibinfo{year}{2016}).

\bibitem[{\citenamefont{{Goldreich} and
  {Reisenegger}}(1992)}]{1992ApJ...395..250G}
\bibinfo{author}{\bibfnamefont{P.}~\bibnamefont{{Goldreich}}} \bibnamefont{and}
  \bibinfo{author}{\bibfnamefont{A.}~\bibnamefont{{Reisenegger}}},
  \bibinfo{journal}{Astrophys. J.} \textbf{\bibinfo{volume}{395}},
  \bibinfo{pages}{250} (\bibinfo{year}{1992}).

\bibitem[{\citenamefont{Comer and Joynt}(2003)}]{Comer:2002dm}
\bibinfo{author}{\bibfnamefont{G.}~\bibnamefont{Comer}} \bibnamefont{and}
  \bibinfo{author}{\bibfnamefont{R.}~\bibnamefont{Joynt}},
  \bibinfo{journal}{Phys.Rev.} \textbf{\bibinfo{volume}{D68}},
  \bibinfo{pages}{023002} (\bibinfo{year}{2003}), \eprint{gr-qc/0212083}.

\bibitem[{\citenamefont{Gusakov et~al.}(2009)\citenamefont{Gusakov, Kantor, and
  Haensel}}]{Gusakov:2009kc}
\bibinfo{author}{\bibfnamefont{M.~E.} \bibnamefont{Gusakov}},
  \bibinfo{author}{\bibfnamefont{E.~M.} \bibnamefont{Kantor}},
  \bibnamefont{and} \bibinfo{author}{\bibfnamefont{P.}~\bibnamefont{Haensel}},
  \bibinfo{journal}{Phys. Rev.} \textbf{\bibinfo{volume}{C79}},
  \bibinfo{pages}{055806} (\bibinfo{year}{2009}), \eprint{0904.3467}.

\bibitem[{\citenamefont{Glampedakis et~al.}(2011)\citenamefont{Glampedakis,
  Jones, and Samuelsson}}]{Glampedakis:2010ec}
\bibinfo{author}{\bibfnamefont{K.}~\bibnamefont{Glampedakis}},
  \bibinfo{author}{\bibfnamefont{D.~I.} \bibnamefont{Jones}}, \bibnamefont{and}
  \bibinfo{author}{\bibfnamefont{L.}~\bibnamefont{Samuelsson}},
  \bibinfo{journal}{Mon. Not. Roy. Astron. Soc.}
  \textbf{\bibinfo{volume}{413}}, \bibinfo{pages}{2021} (\bibinfo{year}{2011}),
  \eprint{1010.1153}.

\bibitem[{\citenamefont{Chamel and Haensel}(2008)}]{Chamel:2008ca}
\bibinfo{author}{\bibfnamefont{N.}~\bibnamefont{Chamel}} \bibnamefont{and}
  \bibinfo{author}{\bibfnamefont{P.}~\bibnamefont{Haensel}},
  \bibinfo{journal}{Living Rev. Rel.} \textbf{\bibinfo{volume}{11}},
  \bibinfo{pages}{10} (\bibinfo{year}{2008}), \eprint{0812.3955}.

\bibitem[{\citenamefont{Schmitt and Shternin}(2018)}]{Schmitt:2017efp}
\bibinfo{author}{\bibfnamefont{A.}~\bibnamefont{Schmitt}} \bibnamefont{and}
  \bibinfo{author}{\bibfnamefont{P.}~\bibnamefont{Shternin}},
  \bibinfo{journal}{Astrophys. Space Sci. Libr.}
  \textbf{\bibinfo{volume}{457}}, \bibinfo{pages}{455} (\bibinfo{year}{2018}),
  \eprint{1711.06520}.

\bibitem[{\citenamefont{Wu and Niu}(2001)}]{PhysRevA.64.061603}
\bibinfo{author}{\bibfnamefont{B.}~\bibnamefont{Wu}} \bibnamefont{and}
  \bibinfo{author}{\bibfnamefont{Q.}~\bibnamefont{Niu}},
  \bibinfo{journal}{Phys. Rev. A} \textbf{\bibinfo{volume}{64}},
  \bibinfo{pages}{061603} (\bibinfo{year}{2001}), \eprint{cond-mat/0009455}.

\bibitem[{\citenamefont{Tisza}(1938)}]{tisza38}
\bibinfo{author}{\bibfnamefont{L.}~\bibnamefont{Tisza}},
  \bibinfo{journal}{Nature} \textbf{\bibinfo{volume}{141}},
  \bibinfo{pages}{913} (\bibinfo{year}{1938}).

\bibitem[{\citenamefont{Landau}(1941)}]{landau41}
\bibinfo{author}{\bibfnamefont{L.}~\bibnamefont{Landau}},
  \bibinfo{journal}{Phys. Rev.} \textbf{\bibinfo{volume}{60}},
  \bibinfo{pages}{356} (\bibinfo{year}{1941}).

\bibitem[{\citenamefont{Alford et~al.}(2013)\citenamefont{Alford, Mallavarapu,
  Schmitt, and Stetina}}]{Alford:2012vn}
\bibinfo{author}{\bibfnamefont{M.~G.} \bibnamefont{Alford}},
  \bibinfo{author}{\bibfnamefont{S.~K.} \bibnamefont{Mallavarapu}},
  \bibinfo{author}{\bibfnamefont{A.}~\bibnamefont{Schmitt}}, \bibnamefont{and}
  \bibinfo{author}{\bibfnamefont{S.}~\bibnamefont{Stetina}},
  \bibinfo{journal}{Phys. Rev.} \textbf{\bibinfo{volume}{D87}},
  \bibinfo{pages}{065001} (\bibinfo{year}{2013}), \eprint{1212.0670}.

\bibitem[{\citenamefont{Kurkela et~al.}(2018)\citenamefont{Kurkela,
  Mukhopadhyay, Preis, Rebhan, and Soloviev}}]{Kurkela:2018dku}
\bibinfo{author}{\bibfnamefont{A.}~\bibnamefont{Kurkela}},
  \bibinfo{author}{\bibfnamefont{A.}~\bibnamefont{Mukhopadhyay}},
  \bibinfo{author}{\bibfnamefont{F.}~\bibnamefont{Preis}},
  \bibinfo{author}{\bibfnamefont{A.}~\bibnamefont{Rebhan}}, \bibnamefont{and}
  \bibinfo{author}{\bibfnamefont{A.}~\bibnamefont{Soloviev}},
  \bibinfo{journal}{JHEP} \textbf{\bibinfo{volume}{08}}, \bibinfo{pages}{054}
  (\bibinfo{year}{2018}), \eprint{1805.05213}.

\bibitem[{\citenamefont{Leung et~al.}(2011)\citenamefont{Leung, Chu, and
  Lin}}]{Leung:2011zz}
\bibinfo{author}{\bibfnamefont{S.~C.} \bibnamefont{Leung}},
  \bibinfo{author}{\bibfnamefont{M.~C.} \bibnamefont{Chu}}, \bibnamefont{and}
  \bibinfo{author}{\bibfnamefont{L.~M.} \bibnamefont{Lin}},
  \bibinfo{journal}{Phys. Rev.} \textbf{\bibinfo{volume}{D84}},
  \bibinfo{pages}{107301} (\bibinfo{year}{2011}), \eprint{1111.1787}.

\bibitem[{\citenamefont{Xiang et~al.}(2014)\citenamefont{Xiang, Jiang, Zhang,
  and Yang}}]{Xiang:2013xwa}
\bibinfo{author}{\bibfnamefont{Q.-F.} \bibnamefont{Xiang}},
  \bibinfo{author}{\bibfnamefont{W.-Z.} \bibnamefont{Jiang}},
  \bibinfo{author}{\bibfnamefont{D.-R.} \bibnamefont{Zhang}}, \bibnamefont{and}
  \bibinfo{author}{\bibfnamefont{R.-Y.} \bibnamefont{Yang}},
  \bibinfo{journal}{Phys. Rev.} \textbf{\bibinfo{volume}{C89}},
  \bibinfo{pages}{025803} (\bibinfo{year}{2014}), \eprint{1305.7354}.

\bibitem[{\citenamefont{Mukhopadhyay and
  Schaffner-Bielich}(2016)}]{Mukhopadhyay:2015xhs}
\bibinfo{author}{\bibfnamefont{P.}~\bibnamefont{Mukhopadhyay}}
  \bibnamefont{and}
  \bibinfo{author}{\bibfnamefont{J.}~\bibnamefont{Schaffner-Bielich}},
  \bibinfo{journal}{Phys. Rev.} \textbf{\bibinfo{volume}{D93}},
  \bibinfo{pages}{083009} (\bibinfo{year}{2016}), \eprint{1511.00238}.

\bibitem[{\citenamefont{Buneman}(1959)}]{Buneman:1959zz}
\bibinfo{author}{\bibfnamefont{O.}~\bibnamefont{Buneman}},
  \bibinfo{journal}{Phys.Rev.} \textbf{\bibinfo{volume}{115}},
  \bibinfo{pages}{503} (\bibinfo{year}{1959}).

\bibitem[{\citenamefont{{Farley}}(1963)}]{1963PhRvL..10..279F}
\bibinfo{author}{\bibfnamefont{D.~T.} \bibnamefont{{Farley}}},
  \bibinfo{journal}{Physical Review Letters} \textbf{\bibinfo{volume}{10}},
  \bibinfo{pages}{279} (\bibinfo{year}{1963}).

\bibitem[{\citenamefont{{Anderson} et~al.}(2001)\citenamefont{{Anderson},
  {Fedele}, and {Lisak}}}]{2001AmJPh..69.1262A}
\bibinfo{author}{\bibfnamefont{D.}~\bibnamefont{{Anderson}}},
  \bibinfo{author}{\bibfnamefont{R.}~\bibnamefont{{Fedele}}}, \bibnamefont{and}
  \bibinfo{author}{\bibfnamefont{M.}~\bibnamefont{{Lisak}}},
  \bibinfo{journal}{American Journal of Physics} \textbf{\bibinfo{volume}{69}},
  \bibinfo{pages}{1262} (\bibinfo{year}{2001}).

\bibitem[{\citenamefont{Livescu et~al.}(2011)\citenamefont{Livescu, Wei, and
  Petersen}}]{Livescu_2011}
\bibinfo{author}{\bibfnamefont{D.}~\bibnamefont{Livescu}},
  \bibinfo{author}{\bibfnamefont{T.}~\bibnamefont{Wei}}, \bibnamefont{and}
  \bibinfo{author}{\bibfnamefont{M.~R.} \bibnamefont{Petersen}},
  \bibinfo{journal}{Journal of Physics: Conference Series}
  \textbf{\bibinfo{volume}{318}}, \bibinfo{pages}{082007}
  (\bibinfo{year}{2011}).

\bibitem[{\citenamefont{{Andersson} et~al.}(2004)\citenamefont{{Andersson},
  {Comer}, and {Prix}}}]{2004MNRAS.354..101A}
\bibinfo{author}{\bibfnamefont{N.}~\bibnamefont{{Andersson}}},
  \bibinfo{author}{\bibfnamefont{G.~L.} \bibnamefont{{Comer}}},
  \bibnamefont{and} \bibinfo{author}{\bibfnamefont{R.}~\bibnamefont{{Prix}}},
  \bibinfo{journal}{Mon.Not.Roy.Astron.Soc.} \textbf{\bibinfo{volume}{354}},
  \bibinfo{pages}{101} (\bibinfo{year}{2004}).

\bibitem[{\citenamefont{Haber et~al.}(2016)\citenamefont{Haber, Schmitt, and
  Stetina}}]{Haber:2015exa}
\bibinfo{author}{\bibfnamefont{A.}~\bibnamefont{Haber}},
  \bibinfo{author}{\bibfnamefont{A.}~\bibnamefont{Schmitt}}, \bibnamefont{and}
  \bibinfo{author}{\bibfnamefont{S.}~\bibnamefont{Stetina}},
  \bibinfo{journal}{Phys. Rev.} \textbf{\bibinfo{volume}{D93}},
  \bibinfo{pages}{025011} (\bibinfo{year}{2016}), \eprint{1510.01982}.

\bibitem[{\citenamefont{{Ruostekoski} and
  {Dutton}}(2007)}]{2007PhRvA..76f3607R}
\bibinfo{author}{\bibfnamefont{J.}~\bibnamefont{{Ruostekoski}}}
  \bibnamefont{and} \bibinfo{author}{\bibfnamefont{Z.}~\bibnamefont{{Dutton}}},
  \bibinfo{journal}{\pra} \textbf{\bibinfo{volume}{76}}, \bibinfo{eid}{063607}
  (\bibinfo{year}{2007}), \eprint{0707.2571}.

\bibitem[{\citenamefont{Yu et~al.}(2018)\citenamefont{Yu, Chai, and
  Xue}}]{YU20181231}
\bibinfo{author}{\bibfnamefont{Z.-F.} \bibnamefont{Yu}},
  \bibinfo{author}{\bibfnamefont{X.-D.} \bibnamefont{Chai}}, \bibnamefont{and}
  \bibinfo{author}{\bibfnamefont{J.-K.} \bibnamefont{Xue}},
  \bibinfo{journal}{Physics Letters A} \textbf{\bibinfo{volume}{382}},
  \bibinfo{pages}{1231 } (\bibinfo{year}{2018}).

\bibitem[{\citenamefont{{Friedman} and {Schutz}}(1978)}]{1978ApJ...221..937F}
\bibinfo{author}{\bibfnamefont{J.~L.} \bibnamefont{{Friedman}}}
  \bibnamefont{and} \bibinfo{author}{\bibfnamefont{B.~F.}
  \bibnamefont{{Schutz}}}, \bibinfo{journal}{Astrophys. J.}
  \textbf{\bibinfo{volume}{221}}, \bibinfo{pages}{937} (\bibinfo{year}{1978}).

\bibitem[{\citenamefont{Khalatnikov}(1989)}]{khala}
\bibinfo{author}{\bibfnamefont{I.}~\bibnamefont{Khalatnikov}},
  \emph{\bibinfo{title}{An Introduction to the Theory of Superfluidity}}
  (\bibinfo{publisher}{Addison-Wesley}, \bibinfo{address}{New York},
  \bibinfo{year}{1989}).

\bibitem[{\citenamefont{Mannarelli and Manuel}(2010)}]{Mannarelli:2009ia}
\bibinfo{author}{\bibfnamefont{M.}~\bibnamefont{Mannarelli}} \bibnamefont{and}
  \bibinfo{author}{\bibfnamefont{C.}~\bibnamefont{Manuel}},
  \bibinfo{journal}{Phys. Rev.} \textbf{\bibinfo{volume}{D81}},
  \bibinfo{pages}{043002} (\bibinfo{year}{2010}), \eprint{0909.4486}.

\bibitem[{\citenamefont{Hiscock and Lindblom}(1985)}]{Hiscock:1985zz}
\bibinfo{author}{\bibfnamefont{W.~A.} \bibnamefont{Hiscock}} \bibnamefont{and}
  \bibinfo{author}{\bibfnamefont{L.}~\bibnamefont{Lindblom}},
  \bibinfo{journal}{Phys. Rev.} \textbf{\bibinfo{volume}{D31}},
  \bibinfo{pages}{725} (\bibinfo{year}{1985}).

\bibitem[{\citenamefont{Hiscock and Lindblom}(1987)}]{Hiscock:1987zz}
\bibinfo{author}{\bibfnamefont{W.~A.} \bibnamefont{Hiscock}} \bibnamefont{and}
  \bibinfo{author}{\bibfnamefont{L.}~\bibnamefont{Lindblom}},
  \bibinfo{journal}{Phys. Rev.} \textbf{\bibinfo{volume}{D35}},
  \bibinfo{pages}{3723} (\bibinfo{year}{1987}).

\bibitem[{\citenamefont{Kovtun}(2019)}]{Kovtun:2019hdm}
\bibinfo{author}{\bibfnamefont{P.}~\bibnamefont{Kovtun}},
  \bibinfo{journal}{JHEP} \textbf{\bibinfo{volume}{10}}, \bibinfo{pages}{034}
  (\bibinfo{year}{2019}), \eprint{1907.08191}.

\bibitem[{\citenamefont{Bemfica et~al.}(2019)\citenamefont{Bemfica, Disconzi,
  and Noronha}}]{Bemfica:2019knx}
\bibinfo{author}{\bibfnamefont{F.~S.} \bibnamefont{Bemfica}},
  \bibinfo{author}{\bibfnamefont{M.~M.} \bibnamefont{Disconzi}},
  \bibnamefont{and} \bibinfo{author}{\bibfnamefont{J.}~\bibnamefont{Noronha}}
  (\bibinfo{year}{2019}), \eprint{1907.12695}.

\bibitem[{\citenamefont{Amado et~al.}(2014)\citenamefont{Amado, Areán,
  Jiménez-Alba, Landsteiner, Melgar, and Salazar~Landea}}]{Amado:2013aea}
\bibinfo{author}{\bibfnamefont{I.}~\bibnamefont{Amado}},
  \bibinfo{author}{\bibfnamefont{D.}~\bibnamefont{Areán}},
  \bibinfo{author}{\bibfnamefont{A.}~\bibnamefont{Jiménez-Alba}},
  \bibinfo{author}{\bibfnamefont{K.}~\bibnamefont{Landsteiner}},
  \bibinfo{author}{\bibfnamefont{L.}~\bibnamefont{Melgar}}, \bibnamefont{and}
  \bibinfo{author}{\bibfnamefont{I.}~\bibnamefont{Salazar~Landea}},
  \bibinfo{journal}{JHEP} \textbf{\bibinfo{volume}{02}}, \bibinfo{pages}{063}
  (\bibinfo{year}{2014}), \eprint{1307.8100}.

\bibitem[{\citenamefont{Alford et~al.}(2014)\citenamefont{Alford, Mallavarapu,
  Schmitt, and Stetina}}]{Alford:2013koa}
\bibinfo{author}{\bibfnamefont{M.~G.} \bibnamefont{Alford}},
  \bibinfo{author}{\bibfnamefont{S.~K.} \bibnamefont{Mallavarapu}},
  \bibinfo{author}{\bibfnamefont{A.}~\bibnamefont{Schmitt}}, \bibnamefont{and}
  \bibinfo{author}{\bibfnamefont{S.}~\bibnamefont{Stetina}},
  \bibinfo{journal}{Phys. Rev.} \textbf{\bibinfo{volume}{D89}},
  \bibinfo{pages}{085005} (\bibinfo{year}{2014}), \eprint{1310.5953}.

\bibitem[{\citenamefont{Schmitt}(2014)}]{Schmitt:2013nva}
\bibinfo{author}{\bibfnamefont{A.}~\bibnamefont{Schmitt}},
  \bibinfo{journal}{Phys. Rev.} \textbf{\bibinfo{volume}{D89}},
  \bibinfo{pages}{065024} (\bibinfo{year}{2014}), \eprint{1312.5993}.

\bibitem[{\citenamefont{Peralta et~al.}(2006)\citenamefont{Peralta, Melatos,
  Giacobello, and Ooi}}]{Peralta:2006um}
\bibinfo{author}{\bibfnamefont{C.}~\bibnamefont{Peralta}},
  \bibinfo{author}{\bibfnamefont{A.}~\bibnamefont{Melatos}},
  \bibinfo{author}{\bibfnamefont{M.}~\bibnamefont{Giacobello}},
  \bibnamefont{and} \bibinfo{author}{\bibfnamefont{A.}~\bibnamefont{Ooi}},
  \bibinfo{journal}{Astrophys. J.} \textbf{\bibinfo{volume}{651}},
  \bibinfo{pages}{1079} (\bibinfo{year}{2006}), \eprint{astro-ph/0607161}.

\bibitem[{\citenamefont{Chamel et~al.}(2012)\citenamefont{Chamel, Pearson, and
  Goriely}}]{Chamel:2012pk}
\bibinfo{author}{\bibfnamefont{N.}~\bibnamefont{Chamel}},
  \bibinfo{author}{\bibfnamefont{J.~M.} \bibnamefont{Pearson}},
  \bibnamefont{and} \bibinfo{author}{\bibfnamefont{S.}~\bibnamefont{Goriely}},
  \bibinfo{journal}{ASP Conf. Ser.} \textbf{\bibinfo{volume}{466}},
  \bibinfo{pages}{203} (\bibinfo{year}{2012}), \eprint{1206.6926}.

\bibitem[{\citenamefont{Andersson et~al.}(2012)\citenamefont{Andersson,
  Glampedakis, Ho, and Espinoza}}]{Andersson:2012iu}
\bibinfo{author}{\bibfnamefont{N.}~\bibnamefont{Andersson}},
  \bibinfo{author}{\bibfnamefont{K.}~\bibnamefont{Glampedakis}},
  \bibinfo{author}{\bibfnamefont{W.~C.~G.} \bibnamefont{Ho}}, \bibnamefont{and}
  \bibinfo{author}{\bibfnamefont{C.~M.} \bibnamefont{Espinoza}},
  \bibinfo{journal}{Phys. Rev. Lett.} \textbf{\bibinfo{volume}{109}},
  \bibinfo{pages}{241103} (\bibinfo{year}{2012}), \eprint{1207.0633}.

\bibitem[{\citenamefont{Haskell and Melatos}(2015)}]{Haskell:2015jra}
\bibinfo{author}{\bibfnamefont{B.}~\bibnamefont{Haskell}} \bibnamefont{and}
  \bibinfo{author}{\bibfnamefont{A.}~\bibnamefont{Melatos}},
  \bibinfo{journal}{Int. J. Mod. Phys.} \textbf{\bibinfo{volume}{D24}},
  \bibinfo{pages}{1530008} (\bibinfo{year}{2015}), \eprint{1502.07062}.

\bibitem[{\citenamefont{{Hunter}}(1977)}]{1977ApJ...213..497H}
\bibinfo{author}{\bibfnamefont{C.}~\bibnamefont{{Hunter}}},
  \bibinfo{journal}{\apj} \textbf{\bibinfo{volume}{213}}, \bibinfo{pages}{497}
  (\bibinfo{year}{1977}).

\bibitem[{\citenamefont{Friedman and Schutz}(1978)}]{Friedman:1978hf}
\bibinfo{author}{\bibfnamefont{J.~L.} \bibnamefont{Friedman}} \bibnamefont{and}
  \bibinfo{author}{\bibfnamefont{B.~F.} \bibnamefont{Schutz}},
  \bibinfo{journal}{Astrophys. J.} \textbf{\bibinfo{volume}{222}},
  \bibinfo{pages}{281} (\bibinfo{year}{1978}).

\bibitem[{\citenamefont{Chandrasekhar}(1970)}]{PhysRevLett.24.611}
\bibinfo{author}{\bibfnamefont{S.}~\bibnamefont{Chandrasekhar}},
  \bibinfo{journal}{Phys. Rev. Lett.} \textbf{\bibinfo{volume}{24}},
  \bibinfo{pages}{611} (\bibinfo{year}{1970}).

\bibitem[{\citenamefont{{Ipser} and {Lindblom}}(1991)}]{1991ApJ...373..213I}
\bibinfo{author}{\bibfnamefont{J.~R.} \bibnamefont{{Ipser}}} \bibnamefont{and}
  \bibinfo{author}{\bibfnamefont{L.}~\bibnamefont{{Lindblom}}},
  \bibinfo{journal}{Astrophys. J.} \textbf{\bibinfo{volume}{373}},
  \bibinfo{pages}{213} (\bibinfo{year}{1991}).

\bibitem[{\citenamefont{Gaertig et~al.}(2011)\citenamefont{Gaertig,
  Glampedakis, Kokkotas, and Zink}}]{Gaertig:2011bm}
\bibinfo{author}{\bibfnamefont{E.}~\bibnamefont{Gaertig}},
  \bibinfo{author}{\bibfnamefont{K.}~\bibnamefont{Glampedakis}},
  \bibinfo{author}{\bibfnamefont{K.~D.} \bibnamefont{Kokkotas}},
  \bibnamefont{and} \bibinfo{author}{\bibfnamefont{B.}~\bibnamefont{Zink}},
  \bibinfo{journal}{Phys. Rev. Lett.} \textbf{\bibinfo{volume}{107}},
  \bibinfo{pages}{101102} (\bibinfo{year}{2011}), \eprint{1106.5512}.

\bibitem[{\citenamefont{Andersson}(1998)}]{Andersson:1997xt}
\bibinfo{author}{\bibfnamefont{N.}~\bibnamefont{Andersson}},
  \bibinfo{journal}{Astrophys. J.} \textbf{\bibinfo{volume}{502}},
  \bibinfo{pages}{708} (\bibinfo{year}{1998}), \eprint{gr-qc/9706075}.

\bibitem[{\citenamefont{Glampedakis and Gualtieri}(2018)}]{Glampedakis:2017nqy}
\bibinfo{author}{\bibfnamefont{K.}~\bibnamefont{Glampedakis}} \bibnamefont{and}
  \bibinfo{author}{\bibfnamefont{L.}~\bibnamefont{Gualtieri}},
  \bibinfo{journal}{Astrophys. Space Sci. Libr.}
  \textbf{\bibinfo{volume}{457}}, \bibinfo{pages}{673} (\bibinfo{year}{2018}),
  \eprint{1709.07049}.

\bibitem[{\citenamefont{Kapusta et~al.}(2012)\citenamefont{Kapusta, Muller, and
  Stephanov}}]{Kapusta:2011gt}
\bibinfo{author}{\bibfnamefont{J.~I.} \bibnamefont{Kapusta}},
  \bibinfo{author}{\bibfnamefont{B.}~\bibnamefont{Muller}}, \bibnamefont{and}
  \bibinfo{author}{\bibfnamefont{M.}~\bibnamefont{Stephanov}},
  \bibinfo{journal}{Phys. Rev.} \textbf{\bibinfo{volume}{C85}},
  \bibinfo{pages}{054906} (\bibinfo{year}{2012}), \eprint{1112.6405}.

\bibitem[{\citenamefont{Kovtun}(2012)}]{Kovtun:2012rj}
\bibinfo{author}{\bibfnamefont{P.}~\bibnamefont{Kovtun}}, \bibinfo{journal}{J.
  Phys.} \textbf{\bibinfo{volume}{A45}}, \bibinfo{pages}{473001}
  (\bibinfo{year}{2012}), \eprint{1205.5040}.

\bibitem[{\citenamefont{Strickland}(2014)}]{Strickland:2014pga}
\bibinfo{author}{\bibfnamefont{M.}~\bibnamefont{Strickland}},
  \bibinfo{journal}{Acta Phys. Polon.} \textbf{\bibinfo{volume}{B45}},
  \bibinfo{pages}{2355} (\bibinfo{year}{2014}), \eprint{1410.5786}.

\bibitem[{\citenamefont{Eckart}(1940)}]{PhysRev.58.919}
\bibinfo{author}{\bibfnamefont{C.}~\bibnamefont{Eckart}},
  \bibinfo{journal}{Phys. Rev.} \textbf{\bibinfo{volume}{58}},
  \bibinfo{pages}{919} (\bibinfo{year}{1940}).

\bibitem[{\citenamefont{Landau and Lifshitz}(1987)}]{Landau:Fluid}
\bibinfo{author}{\bibfnamefont{L.~D.} \bibnamefont{Landau}} \bibnamefont{and}
  \bibinfo{author}{\bibfnamefont{E.~M.} \bibnamefont{Lifshitz}},
  \emph{\bibinfo{title}{Fluid Mechanics}} (\bibinfo{publisher}{Pergamon},
  \bibinfo{address}{Oxford}, \bibinfo{year}{1987}), \bibinfo{edition}{2nd} ed.

\bibitem[{\citenamefont{Bitaghsir~Fadafan
  et~al.}(2019)\citenamefont{Bitaghsir~Fadafan, Kazemian, and
  Schmitt}}]{BitaghsirFadafan:2018uzs}
\bibinfo{author}{\bibfnamefont{K.}~\bibnamefont{Bitaghsir~Fadafan}},
  \bibinfo{author}{\bibfnamefont{F.}~\bibnamefont{Kazemian}}, \bibnamefont{and}
  \bibinfo{author}{\bibfnamefont{A.}~\bibnamefont{Schmitt}},
  \bibinfo{journal}{JHEP} \textbf{\bibinfo{volume}{03}}, \bibinfo{pages}{183}
  (\bibinfo{year}{2019}), \eprint{1811.08698}.

\bibitem[{\citenamefont{Carter and Langlois}(1995)}]{Carter:1995if}
\bibinfo{author}{\bibfnamefont{B.}~\bibnamefont{Carter}} \bibnamefont{and}
  \bibinfo{author}{\bibfnamefont{D.}~\bibnamefont{Langlois}},
  \bibinfo{journal}{Phys. Rev.} \textbf{\bibinfo{volume}{D51}},
  \bibinfo{pages}{5855} (\bibinfo{year}{1995}), \eprint{hep-th/9507058}.

\bibitem[{\citenamefont{Andersson and Comer}(2007)}]{Andersson:2006nr}
\bibinfo{author}{\bibfnamefont{N.}~\bibnamefont{Andersson}} \bibnamefont{and}
  \bibinfo{author}{\bibfnamefont{G.~L.} \bibnamefont{Comer}},
  \bibinfo{journal}{Living Rev. Rel.} \textbf{\bibinfo{volume}{10}},
  \bibinfo{pages}{1} (\bibinfo{year}{2007}), \eprint{gr-qc/0605010}.

\bibitem[{\citenamefont{{Andreev} and {Bashkin}}(1975)}]{1976JETP...42..164A}
\bibinfo{author}{\bibfnamefont{A.~F.} \bibnamefont{{Andreev}}}
  \bibnamefont{and} \bibinfo{author}{\bibfnamefont{E.~P.}
  \bibnamefont{{Bashkin}}}, \bibinfo{journal}{Zh. Eksp. Teor. Fiz.}
  \textbf{\bibinfo{volume}{69}}, \bibinfo{pages}{319} (\bibinfo{year}{1975}),
  \bibinfo{note}{[Sov.\ Phys.\ JETP {\bf 42}, 164 (1976)]}.

\bibitem[{\citenamefont{Atkins}(1959)}]{PhysRev.113.962}
\bibinfo{author}{\bibfnamefont{K.~R.} \bibnamefont{Atkins}},
  \bibinfo{journal}{Phys. Rev.} \textbf{\bibinfo{volume}{113}},
  \bibinfo{pages}{962} (\bibinfo{year}{1959}).

\bibitem[{\citenamefont{Yarom}(2009)}]{Yarom:2009uq}
\bibinfo{author}{\bibfnamefont{A.}~\bibnamefont{Yarom}},
  \bibinfo{journal}{JHEP} \textbf{\bibinfo{volume}{07}}, \bibinfo{pages}{070}
  (\bibinfo{year}{2009}), \eprint{0903.1353}.

\bibitem[{\citenamefont{Schmitt}(2015)}]{Schmitt:2014eka}
\bibinfo{author}{\bibfnamefont{A.}~\bibnamefont{Schmitt}},
  \bibinfo{journal}{Lect. Notes Phys.} \textbf{\bibinfo{volume}{888}},
  \bibinfo{pages}{pp.1} (\bibinfo{year}{2015}), \eprint{1404.1284}.

\end{thebibliography}

\end{document}